\pdfoutput=1
\documentclass[%
 reprint,
superscriptaddress,
bibnotes,
tightenlines,
notitlepage,
longbibliography,
twocolumn,
 amsmath,amssymb,
 aps,
]{revtex4-1}

\usepackage{graphicx}

\usepackage{dcolumn}
\usepackage{bm}
\usepackage[usenames,dvipsnames,svgnames,table]{xcolor}
\usepackage[colorlinks = true,
            linkcolor = Navy,
            urlcolor  = Navy,
            citecolor = Navy,
            anchorcolor = Navy]{hyperref}
\usepackage[caption=false]{subfig}

\usepackage{feynmp-auto}

\usepackage{lineno}
\usepackage{makecell}

\usepackage{booktabs}
\newcolumntype{C}{>{$}c<{$}}
\AtBeginDocument{
\heavyrulewidth=.08em
\lightrulewidth=.05em
\cmidrulewidth=.03em
\belowrulesep=.65ex
\belowbottomsep=0pt
\aboverulesep=.4ex
\abovetopsep=0pt
\cmidrulesep=\doublerulesep
\cmidrulekern=.5em
\defaultaddspace=.5em
}

\usepackage[utf8]{inputenc}

\begin{document}

\title{ New physics and tau $g-2$ using LHC heavy ion collisions }% 

\author{Lydia Beresford}
 \email{lydia.beresford@physics.ox.ac.uk}
 \affiliation{%
 Department of Physics, University of Oxford, Oxford OX1 3RH, UK
}
 \author{Jesse Liu}
 \email{jesseliu@uchicago.edu}
 \affiliation{%
 Department of Physics, University of Oxford, Oxford OX1 3RH, UK
} 
\affiliation{% 
Department of Physics, University of Chicago,
Chicago IL 60637, USA
}

%\date{\today}

\begin{abstract}
The anomalous magnetic moment of the tau lepton $a_\tau = (g_\tau -2)/2$ strikingly evades measurement, but is highly sensitive to new physics such as compositeness or supersymmetry. We propose using ultraperipheral heavy ion collisions at the LHC to probe modified magnetic $\delta a_\tau$ and electric dipole moments $\delta d_\tau$. We introduce a suite of one electron/muon plus track(s) analyses, leveraging the exceptionally clean photon fusion $\gamma\gamma \to \tau\tau$ events to reconstruct both leptonic and hadronic tau decays sensitive to $\delta a_\tau, \delta d_\tau$. Assuming 10\% systematic uncertainties, the current 2~nb$^{-1}$ lead--lead dataset could already provide constraints of $-0.0080 < a_\tau < 0.0046$ at 68\% CL. This surpasses 15 year old lepton collider precision by a factor of three while opening novel avenues to new physics.
\end{abstract}

\maketitle

\section{\label{sec:intro} Introduction}

Precision measurements of electromagnetic couplings are foundational tests of quantum electrodynamics (QED) and powerful probes of beyond the Standard Model (BSM) physics. The electron anomalous magnetic moment $a_e = \frac{1}{2}(g_e-2)$ is among the most precisely known quantities in nature~\cite{PhysRevLett.97.030801,Hanneke:2010au,PhysRevLett.106.080801,Aoyama:2012wj,Parker191}. The muon counterpart $a_\mu$ is measured to $10^{-7}$ precision~\cite{Bennett:2006fi} and reports a $3-4\sigma$ tension from SM predictions~\cite{Aoyama:2012wk,Keshavarzi:2018mgv}. This may indicate new physics~\cite{Martin:2001st,Czarnecki:2001pv,Hagiwara:2011af,Ajaib:2015yma}, to be clarified at Fermilab~\cite{Grange:2015fou} and J--PARC~\cite{Abe:2019thb}. Measuring $a_\ell$ generically tests lepton compositeness~\cite{Silverman:1982ft}, while supersymmetry at energy scales $M_\text{S}$ induces radiative corrections $\delta a_{\ell} \sim m_{\ell}^2 / M_\text{S}^2$ for leptons with mass $m_\ell$~\cite{Martin:2001st}. Thus the tau $\tau$ can be $m_\tau^2/m_\mu^2 \sim 280$ times more sensitive to BSM physics than $a_\mu$.  

However, $a_\tau$ continues to evade measurement because the short tau proper lifetime $\sim 10^{-13}$~s precludes use of spin precession methods~\cite{Bennett:2006fi}.
The most precise single-experiment measurement  $a_\tau^\text{exp}$ is from DELPHI~\cite{Abdallah:2003xd,Tanabashi:2018oca} at the Large Electron Positron Collider (LEP), but is remarkably an order of magnitude away from the theoretical central value $a_{\tau,\, \text{SM}}^\text{pred}$ predicted to $10^{-5}$ precision~\cite{Eidelman:2007sb}
\begin{linenomath*}
\begin{align}
    a_\tau^\text{exp} =  - 0.018\,(17),\quad a_{\tau,\,\text{SM}}^\text{pred} = 0.001\,177\,21\,(5).
\end{align}
\end{linenomath*}
The poor constraints on $a_\tau$ present striking room for BSM physics, especially given other lepton sector tensions~\cite{Dutta:2018fge,Davoudiasl:2018fbb,Bauer:2019gfk,Aaij:2015yra,Abdesselam:2019dgh,Allanach:2015gkd,DiChiara:2017cjq,Biswas:2019twf}, and motivate new experimental strategies.

\begin{figure}
    \centering
    \includegraphics[width=0.3\textwidth]{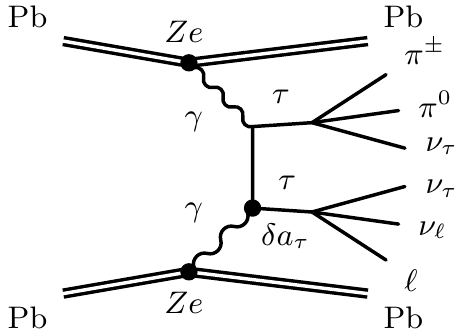}
    \caption{\label{fig:feynGraphs_aa2tautau}Pair production of tau leptons $\tau$ from ultraperipheral lead ion (Pb) collisions in two of the most common decay modes: $\pi^\pm \pi^0 \nu_\tau$ and $\ell \nu_\ell \nu_\tau$. New physics can modify tau--photon couplings affecting the magnetic moment by $\delta a_\tau$. }   
\end{figure}

This Letter proposes a suite of analyses to probe $a_\tau$ using heavy ion beams at the LHC. We leverage ultraperipheral collisions (UPC) where only the electromagnetic fields surrounding lead (Pb) ions interact. Tau pairs are produced from photon fusion $\text{PbPb} \to \text{Pb}(\gamma\gamma \to \tau\tau) \text{Pb}$, illustrated in Fig.~\ref{fig:feynGraphs_aa2tautau}, whose sensitivity to $a_\tau$ was suggested in 1991~\cite{DELAGUILA1991256}. We introduce the strategy crucial for experimental realization and importantly show that the currently recorded dataset could already surpass LEP precision. The LHC cross-section enjoys a $Z^4$ enhancement ($Z=82$ for Pb), with over one million $\gamma\gamma \to \tau\tau$ events produced to date. Existing proposals using lepton beams require future datasets (Belle-II) or proposed facilities (CLIC, LHeC)~\cite{Koksal:2018env,Howard:2018aqy,Koksal:2018xyi,Gutierrez-Rodriguez:2019umw,Fael:2013ij,Eidelman:2016aih,Chen:2018cxt}, while LHC studies focus on high luminosity proton beams~\cite{Samuel1994,Hayreter:2013vna,Atag:2010ja,Hayreter:2013vna,Galon:2016ngp,Fomin:2018ybj,Fu:2019utm}. No LHC analysis of $\gamma\gamma\to\tau\tau$ exists as the taus have insufficient momentum for ATLAS/CMS to record or reconstruct. 

Our proposal overcomes these obstructions in the clean UPC events~\cite{ATLAS-EVENTDISPLAY-2018-009}, enabling selection of individual tracks from tau decays with no other detector activity akin to LEP~\cite{Abdallah:2003xd}.
%,Gau:1997cn,Ackerstaff:1998mt,Tabares:2001xq}. 
We exploit recent advances in low momentum electron/muon identification~\cite{Aaboud:2017leg,Sirunyan:2018iwl,ATLAS-CONF-2019-014} to suppress hadronic backgrounds. We then present a shape analysis sensitive to interfering SM and BSM amplitudes to enhance $a_\tau$ constraints. Our strategy also probes tau electric dipole moments $d_\tau$ induced by charge--parity (CP) violating new physics. This opens key new directions in the heavy ion program amid reviving interest in photon collisions~\cite{Piotrzkowski:2000rx,Albrow:2008pn,deFavereaudeJeneret:2009db} for light-by-light scattering~\cite{dEnterria:2013zqi,Aaboud:2017bwk,Sirunyan:2018fhl,Aad:2019ock}, standard candle processes~\cite{Aaboud:2017oiq,Aaboud:2016dkv,Chatrchyan:2012tv,Khachatryan:2016mud,ATLAS-CONF-2016-025}, and BSM dynamics~\cite{Chapon:2009hh,Fichet:2013gsa,Ellis:2017edi,Knapen:2016moh,Baldenegro:2018hng,Ohnemus:1993qw,Schul:2008sr,HarlandLang:2011ih,Beresford:2018pbt,Harland-Lang:2018hmi,Bruce:2018yzs}.

\section{\label{sec:sim_det} Effective theory \& photon flux}

The anomalous $\tau$ magnetic moment $a_\tau = (g_\tau-2)/2$ is defined by the spin--magnetic Hamiltonian $-\boldsymbol{\mu}_\tau\cdot \mathbf{B} = - (g_\tau e / 2m_\tau)\, \mathbf{S} \cdot \mathbf{B}$. In the Lagrangian formulation of QED, electromagnetic moments arise from the spinor tensor $\sigma^{\mu\nu} = \mathrm{i}[\gamma^\mu, \gamma^\nu]/2$ structure of the fermion current interacting with the photon field strength $F_{\mu\nu}$
\begin{linenomath*}
\begin{align}
    \mathcal{L} = \tfrac{1}{2} \bar{\tau}_\text{L}\sigma^{\mu\nu}  \left(a_\tau \tfrac{e}{2m_\tau} - \mathrm{i} d_\tau \gamma_5 \right) \tau_\text{R} F_{\mu\nu}.
\end{align}
\end{linenomath*}
Here, $\gamma^5$ satisfies the anticommutator $\{\gamma^5, \gamma^\mu\} = 0$, and $\tau_\text{L,R}$ are tau spinors with L,R denoting chirality.  

To introduce BSM modifications of $a_\tau$ and  $d_\tau$, we use SM effective field theory (SMEFT)~\cite{Escribano:1993pq}. This assumes the scale of BSM physics $\Lambda$ is much higher than the probe momentum transfers $q$ i.e.,\ $q^2 \ll \Lambda^2$. At scale $q$,
two dimension-six operators in the Warsaw basis~\cite{Grzadkowski:2010es} modify $a_\tau$ and $d_\tau$ at tree level, as discussed in Ref.~\cite{Escribano:1993pq}
\begin{linenomath*}
\begin{align}
    \mathcal{L}' =  \left(\bar{L}_{\tau}\sigma^{\mu\nu} \tau_R\right) H
    \left[
    \frac{C_{\tau B}}{\Lambda^2} B_{\mu\nu} +
    \frac{C_{\tau W}}{\Lambda^2} W_{\mu\nu}   \right].
    \label{eq:BSMLagrangian}
\end{align}
\end{linenomath*}
Here, $B_{\mu\nu}$ and $W_{\mu\nu}$ are the U(1)$_\text{Y}$ and SU(2)$_\text{L}$ field strengths, $H$ ($L_\tau$) is the Higgs (tau lepton) doublet, and $C_i$ are dimensionless, complex Wilson coefficients. We fix $C_{\tau W} = 0$ to parameterize the two modified moments $(\delta a_\tau, \delta d_\tau)$ using two real parameters $(|C_{\tau B}|/\Lambda^2, \varphi)$~\cite{Eidelman:2016aih}
\begin{linenomath*}
\begin{align}
    \delta a_\tau &= \frac{2m_\tau}{e}\frac{|C_{\tau B}|}{M}\cos \varphi,\quad
    \delta d_\tau = \frac{|C_{\tau B}|}{M} \sin \varphi,
    \label{eq:delta_a_d_tau_defn}
\end{align}
\end{linenomath*}
where $\varphi$ is the complex phase of $C_{\tau B}$, we define $M = \Lambda^2/ (\sqrt{2}v\cos\theta_W)$, $\theta_W$ is the electroweak Weinberg angle, and $v = 246$~GeV. 

In the SM, pair production of electrically charged particles $X$ from photon fusion $\gamma\gamma \to XX$ have analytic cross-sections $\sigma_{\gamma\gamma \to XX}$~\cite{Brodsky:1971ud,PhysRevD.23.1933,HarlandLang:2011ih}. For BSM variations, we employ the flavour-general \textsc{SMEFTsim} package~\cite{Brivio:2017btx}, which implements Eq.~\eqref{eq:BSMLagrangian} in \textsc{FeynRules}~\cite{Alloul:2013bka}. This allows a direct interface with \textsc{MadGraph}~2.6.5~\cite{Alwall:2011uj,Alwall:2014hca} for cross-section calculation and Monte Carlo simulation. To model interference between SM and BSM diagrams, we generate $\gamma\gamma \to \tau\tau$ events with up to two BSM couplings $C_{\tau B}$ in the matrix element.

Turning to the source of photons, these are emitted coherently from electromagnetic fields surrounding the ultrarelativistic ions, which is known as the equivalent photon approximation~\cite{Budnev:1974de}. We follow the \textsc{MadGraph} implementation in Ref.~\cite{dEnterria:2009cwl}, which assumes the LHC exclusive cross-section $\sigma_{\gamma\gamma \to XX}^{(\text{PbPb})}$ is factorized into a convolution of $\sigma_{\gamma\gamma \to XX}$ with the ion photon fluxes $n(x)$ 
\begin{linenomath*}  
\begin{equation}
    \sigma_{\gamma\gamma \to XX}^{(\text{PbPb})} = \int \mathrm{d} x_1 \mathrm{d} x_2 \,
    n(x_1) n(x_2)\,
    \sigma_{\gamma\gamma \to XX},
    \label{eq:xsec_excl_photon}
\end{equation}
\end{linenomath*}
where $x_i = E_i / E_\text{beam}$ is the ratio of the emitted photon energy $E_i$ from ion $i$ with beam energy $E_\text{beam}$. In this factorized prescription, $n(x)$ assumes an analytic form from classical field theory~\cite{Jackson:490457,dEnterria:2009cwl}
\begin{linenomath*}  
\begin{equation}
    n(x) = \frac{2Z^2\alpha}{x\pi} \left\{\bar{x} K_0(\bar{x}) K_1(\bar{x}) - \frac{\bar{x}^2}{2}\left[ K_1^2(\bar{x}) -K_0^2(\bar{x})\right]\right\},
    \label{eq:nphotonflux}
\end{equation}
\end{linenomath*}
where $\bar{x} = xm_N b_\text{min}$, $m_N$ is the nucleon mass $m_N = 0.9315$~GeV, and $Z=82$ for Pb. We set the minimum impact parameter $b_\text{min}$ to be the nuclear radius $b_\text{min} = R_A \simeq 1.2 A^{1/3}~\text{fm} = 6.09 A^{1/3}$~GeV$^{-1}$, where $A = 208$ is the mass number of Pb used at the LHC. We use Ref.~\cite{zhang1996computation} to numerically evaluate the modified Bessel functions of the second kind of first $K_0$ and second $K_1$ order.

We modify \textsc{MadGraph} to use the photon flux  Eq.~\eqref{eq:nphotonflux} for evaluating $\sigma^{(\text{PbPb})}_{\gamma\gamma\to XX}$. This prescription neglects a nonfactorizable term in Eq.~\eqref{eq:xsec_excl_photon}, which models the probability of hadronic interactions $P_{|\mathbf{b}_1 - \mathbf{b}_2|}$, where $\mathbf{b}_i$ is the impact parameter of ion $i$. The \textsc{Superchic}~3.02~\cite{Harland-Lang:2018iur} program includes a complete treatment of $P_{|\mathbf{b}_1 - \mathbf{b}_2|}$, along with nuclear overlap and thickness. Using this, we validate that these simplifications in \textsc{MadGraph} do not majorly impact distributions relevant for this work, namely tau $p_\text{T}$. We generate 3 million $\gamma\gamma\to\tau\tau$ events for each coupling variation at $\sqrt{s_\text{NN}} = 5.02$~TeV. For the SM, we find $\sigma^{(\text{PbPb})}_{\gamma\gamma \to \tau\tau} = 5.7\times 10^5$~nb. To improve generator statistics, we impose $p_\text{T}^{\tau} > 3$ GeV in \textsc{MadGraph}, which has a 21\% efficiency. Due to destructive interference, $\sigma^{(\text{PbPb})}_{\gamma\gamma \to \tau\tau}$ falls to a minimum of $4.7\times 10^5$~nb at $\delta a_\tau \simeq -0.04$ before returning to $5.7\times 10^5$~nb at $\delta a_\tau \simeq -0.09$. Further validation of these effects is in Appendix~\ref{sec:apndx_simulation}. We employ \textsc{Pythia} 8.230~\cite{Sjostrand:2007gs} for decay, shower and hadronization, then use \textsc{Delphes} 3.4.1~\cite{deFavereau:2013fsa} for detector emulation. 

\section{\label{sec:analysis} Proposed analyses}

\begin{figure*}
    \centering
    \includegraphics[width=0.33\textwidth]{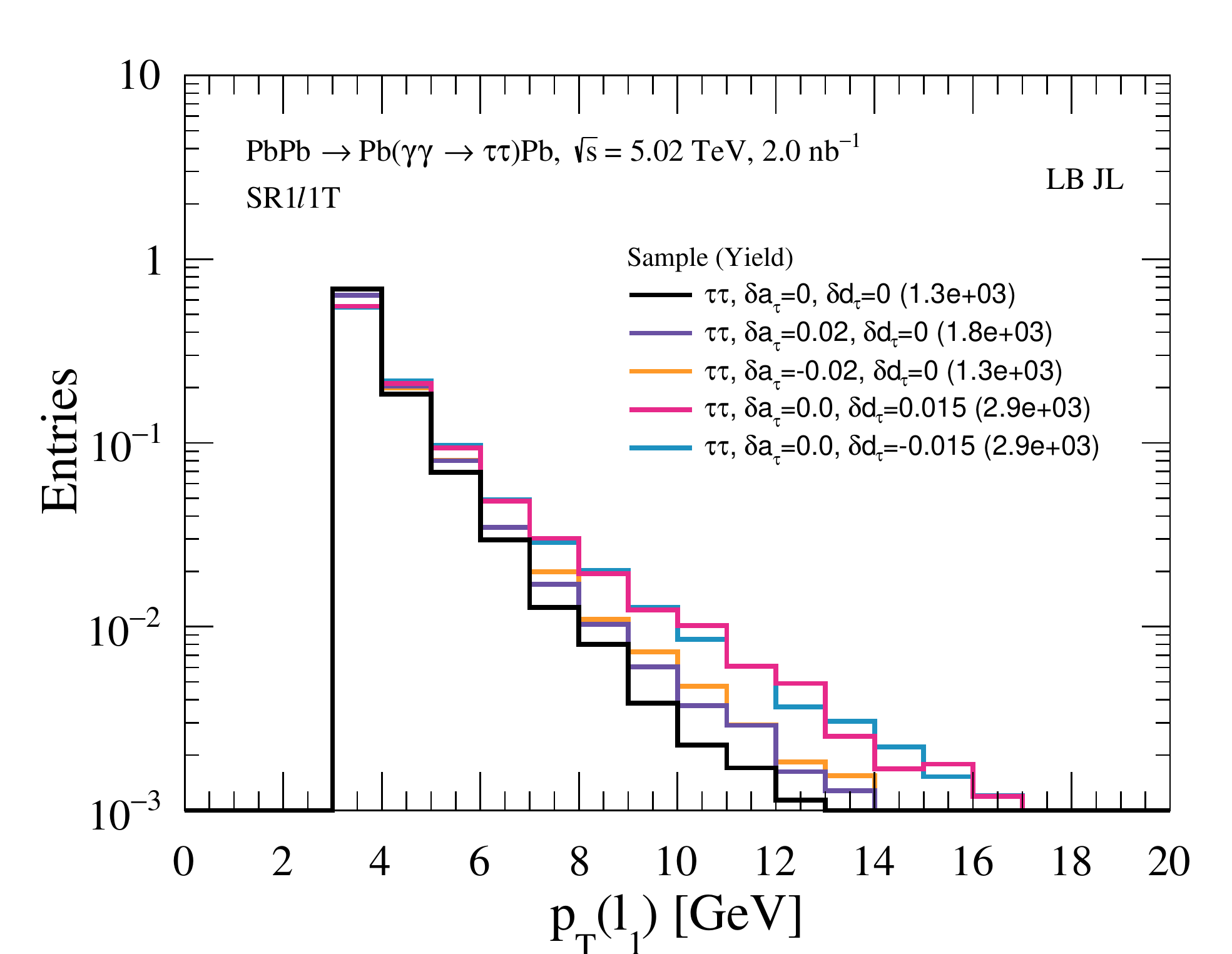}%
    \includegraphics[width=0.33\textwidth]{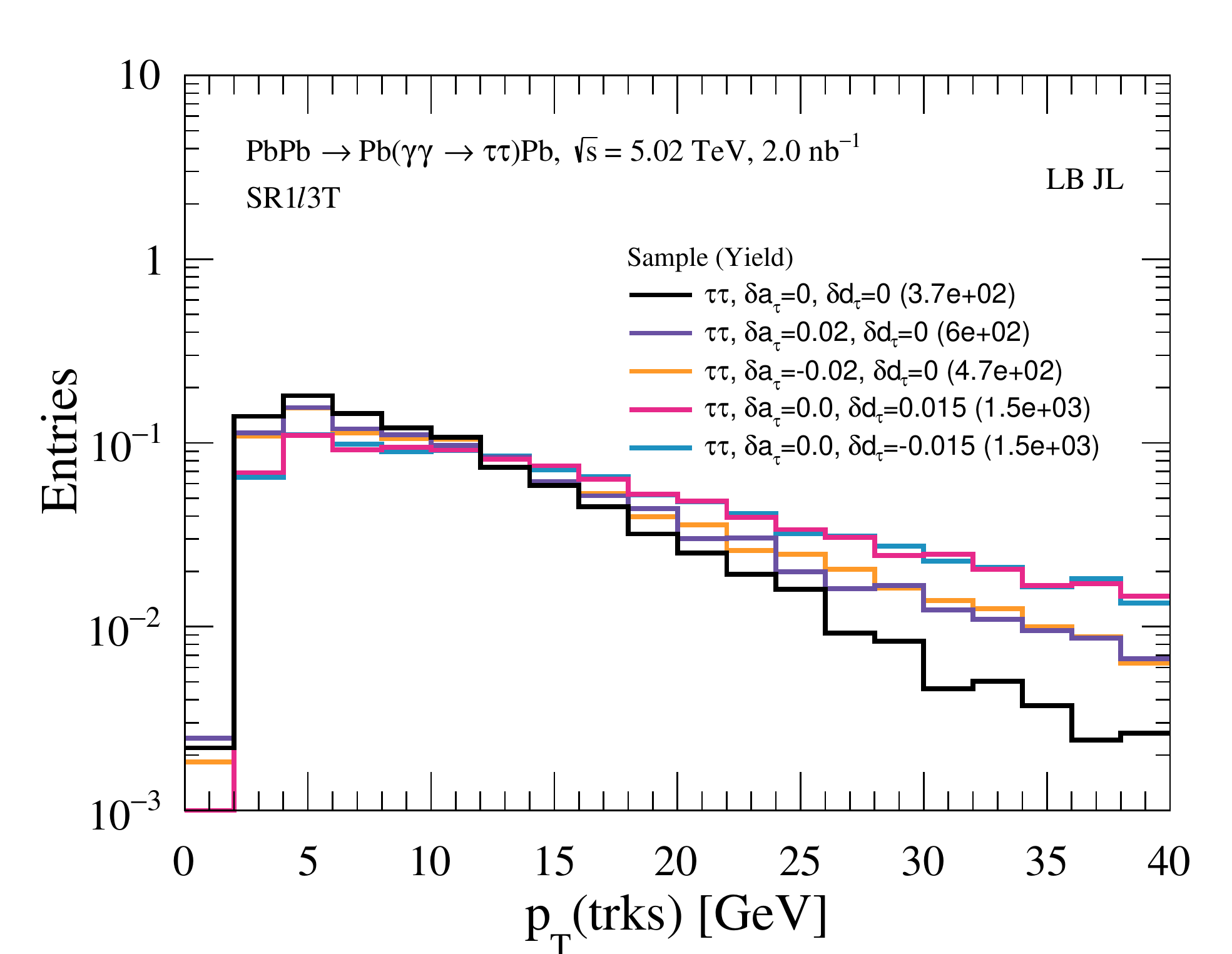}%
    \includegraphics[width=0.33\textwidth]{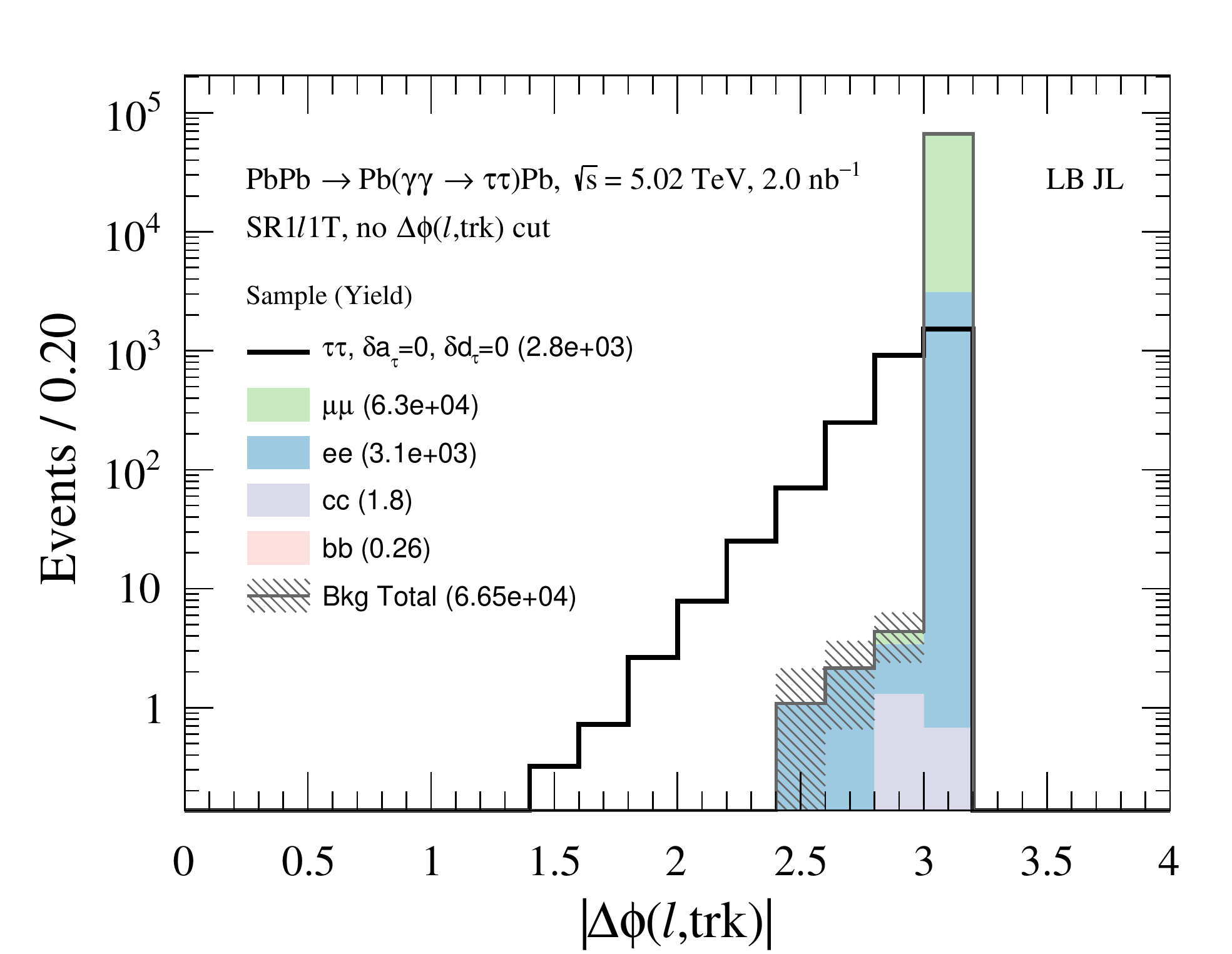}%
    \caption{ 
    Distributions of lepton $p_\text{T}$ in SR$1\ell$1T (left) and the 3-track system $p_\text{T}$ in SR$1\ell$3T (center) for benchmark signals with various $\delta a_\tau$, $\delta d_\tau$ couplings. These are normalized to unit integral to illustrate shape changes with varying $\delta a_\tau, \delta d_\tau$. The lepton--track azimuthal angle $|\Delta \phi(\ell, \text{trk})|$ in SR$1\ell$1T (right) is shown for backgrounds (filled) and signal $\delta a_\tau = \delta d_\tau = 0$ (line), illustrating powerful discrimination against dilepton processes.}
    \label{fig:various_delta_a_tau}
\end{figure*}

To record $\gamma\gamma\to\tau\tau$ events, dedicated UPC triggers are crucial for our proposal. With no other detector activity, the ditau system receives negligible transverse boost and each tau $p_\text{T}$ reaches a few to tens of GeV at most. Taus always decay to a neutrino $\nu_\tau$, which further dilutes the visible momenta, rendering usual hadronic tau triggers $p_\text{T}^\text{$\tau$ jet} \gtrsim 20$~GeV unfeasible~\cite{Aaboud:2016leb,ATLAS-CONF-2017-061}. However, UPC events without pileup enable exceptionally low trigger thresholds by vetoing large sums over calorimeter transverse energy deposits $\sum E_\text{T} < 50$~GeV~\cite{Aad:2019ock}. Other minimum bias triggers are also possible~\cite{ATLAS-CONF-2012-122,ATLAS-CONF-2013-104}. A recent UPC dimuon analysis additionally requires at least one track and no explicit $p_\text{T}$ requirement for the trigger muon~\cite{ATLAS-CONF-2016-025}. The light-by-light observation also considers ultralow $E_\text{T} > 1$~GeV calorimeter cluster thresholds at trigger level~\cite{Aad:2019ock}, which can similarly benefit electrons. 

We design our event selection around two objectives. First, we consider standard objects already deployed by ATLAS/CMS to efficiently reconstruct tau decays with the following branching fractions~\cite{Tanabashi:2018oca}:
\begin{linenomath*}
\begin{align}
    \mathcal{B}(\tau^\pm \to \ell^\pm \nu_\ell \nu_\tau) &= 35\%, \\
    \mathcal{B}(\tau^\pm \to \pi^\pm \nu_\tau + \text{neutral pions}) &= 45.6\%, \\
    \mathcal{B}(\tau^\pm \to \pi^\pm \pi^\mp \pi^\pm\nu_\tau + \text{neutral pions}) &= 19.4\%.
\end{align}
\end{linenomath*}
We develop signal regions (SR) targeting these decays based on expected signal rate and background mitigation strategies. We impose the lowest trigger and reconstruction thresholds $p_\text{T}^{e/\mu} > 4.5/3$~GeV, $|\eta_{e/\mu}| < 2.5/2.4$ supported by ATLAS/CMS~\cite{Aaboud:2017leg,Sirunyan:2018iwl}. Second, we optimize sensitivity to different couplings $\delta a_\tau, \delta d_\tau$, where interfering SM and BSM amplitudes impact tau kinematics, which propagates to e.g.\ lepton $p_\text{T}$. 

\textit{Dilepton analysis}. Requiring two leptons is expected to give the highest signal-to-background $S/B$, with half being different flavor $e\mu$ free of $ee/\mu\mu$ backgrounds. But even using low $p_\text{T}^{e/\mu}$ thresholds, we find insufficient signal yields at 2~nb$^{-1}$ to pursue this further. 
    
\textit{1 lepton + 1 track analysis } (SR$1\ell$1T). This requires exactly 1 lepton and 1 other track that is not `matched' to the lepton (the matched track is the highest $p_\text{T}$ track with $\Delta R(\ell, \text{track}) < 0.02$). Tracks must satisfy the standard requirements $p_\text{T}^\text{track} > 500$~MeV and $|\eta^{\text{track}}| <$ 2.5. This topology targets the high branching ratio of the single charged pion decay mode and background suppression from lepton identification. The track also recovers events failing the dilepton analysis, in which a lepton is too soft to be reconstructed. We divide this SR into two bins $p_\text{T}^{e/\mu} \in [\leq 6], [> 6]$ GeV to exploit shape differences shown in Fig.~\ref{fig:various_delta_a_tau} (left). We require nonplanar lepton--track system $|\Delta \phi(\ell, \text{trk})| < 3$ to suppress back-to-back $ee/\mu\mu$ processes, as demonstrated in Fig.~\ref{fig:various_delta_a_tau} (right). 
We veto invariant masses $m_{\ell, \text{trk}} \not\in [3, 3.2], [9, 11]$~GeV to reject dilepton decays of $J/\psi$ and $\Upsilon$ resonances.
    
\textit{1 lepton + multitrack analysis } (SR$1\ell$2/3T). We augment the previous analysis with 3 non-lepton-matched tracks. This targets the distinctive 3 charged pion decay. We also construct an orthogonal 2 tracks SR to recover misreconstructed 3-pion decays. The non-lepton-matched tracks are used to define the tau candidate as the vectorial sum of the tracks $p_\tau^\text{tracks} = \sum_i p^\text{track}_i$, whose $p_\text{T}$ distribution is shown in Fig.~\ref{fig:various_delta_a_tau} (center) for SR$1\ell$3T. We find removing lepton identification significantly increases hadronic backgrounds.

Leptonic backgrounds are dominated by dielectron/dimuon production $\gamma\gamma \to \ell\ell, \ell \in [e, \mu]$. The single flavor cross-section is sizable $\sigma^{(\text{PbPb})}_{\gamma\gamma \to \ell\ell} = 4.2 \times 10^5$~nb, which includes a generator level $|\eta_\ell| < 2.5$ requirement. The back-to-back leptons are suppressed by the $|\Delta \phi_{\ell\ell}| < 3$ requirement, which we verify by generating 1 million events per flavor. 
Photon radiation from leptons $\ell \to \ell\gamma$ is only expected to modify the tails marginally. Track impact parameters exploiting displaced tau decays could further suppress this background. 
    
Hadronic backgrounds arise from diquark production $\gamma\gamma \to q\bar{q}$ and we generate 1 million events for each of the 5 flavors. For $q \in [u, d, s]$ assuming massless quarks gives a cross-section $\sigma^{(\text{PbPb})}_{\gamma\gamma \to u\bar{u}\,( d\bar{d}, s\bar{s})} = 3.0 \times 10^5$ \,($1.9 \times 10^4$)~nb. Parton showering produces more tracks than tau decays, which we suppress using lepton isolation and requiring no more than 4 tracks at most. For $q \in [c, b]$, heavy flavor $B$ and $D$ mesons undergo semileptonic decays e.g.\ $D\to \pi^0 \ell \nu$. The default \textsc{MadGraph} parameters assume massless charm quarks (which is conservative as a finite mass decreases cross-sections), yielding $\sigma^{(\text{PbPb})}_{\gamma\gamma \to c\bar{c}} = 3.0 \times 10^5$~nb. Bottom quarks assume finite mass resulting in a smaller cross-section $\sigma^{(\text{PbPb})}_{\gamma\gamma \to b\bar{b}} = 1.5 \times 10^3$~nb. The leptonic branching fraction $D\to \pi^0 \ell \nu$ is of order a few percent so is under control, and is further suppressed by isolation.

Smaller potential backgrounds include $\gamma\gamma \to WW$ but the cross-section $\sigma^{(\text{PbPb})}_{\gamma\gamma\to WW} = 14$~pb implies this is safely neglected. Exchange of digluon color singlets (Pomerons) also contributes to diquark backgrounds. These involve strong interactions and as the binding energy per nucleon is very small $\sim 8$~MeV~\cite{dEnterria:2009cwl}, the Pb ions emit more neutrons than QED processes, which can be vetoed by the Zero Degree Calorimeter~\cite{ATLAS:2007aa}. Soft survival for Pomeron exchange is also lower~\cite{dEnterria:2009cwl}, which gives greater activity in the calorimeter and tracker, and are suppressed by our stringent exclusivity requirements. 

Systematic uncertainties require LHC collaborations to reliably quantify, but we discuss expected sources and suggest control strategies. Experimental systematics from current UPC PbPb dimuon measurements have systematics of around 10\%, dominated by luminosity and trigger~\cite{ATLAS-CONF-2016-025}. Systematics from lepton reconstruction are $p_\text{T}^\ell$-dependent and thus sensitive to $\delta a_\tau$. These are most significant at low $p_\text{T}$, but are currently determined in high luminosity proton collisions with challenging backgrounds from fakes~\cite{Aaboud:2019ynx,Aad:2016jkr}, and could be better controlled using clean $\gamma\gamma \to \ell\ell$ events. 

Theoretical uncertainties are expected to be dominated by modeling of the photon flux, nuclear form factors and nucleon dissociation. Fortunately, these initial state effects are independent of QED process and final state. So, experimentalists could use a control sample of $\gamma\gamma\to \ell\ell$ events to constrain these universal nuclear systematics or eliminate them in a ratio analysis with dileptons $\sigma^{(\text{PbPb})}_{\gamma\gamma\to\tau\tau} / \sigma^{(\text{PbPb})}_{\gamma\gamma\to\ell\ell}$. Hadronic backgrounds are susceptible to uncertainties from modeling the parton shower, but are subdominant given $S/B \gg 1$ in our analyses. 
\section{\label{sec:sensitivity} Results \& discussion}

We now estimate the sensitivity of our analyses to modified tau moments $\delta a_\tau, \delta d_\tau$.  Assuming the observed data correspond to the SM expectation, we calculate
\begin{align}
    \chi^2 = \frac{(S_\text{SM+BSM} - S_\text{SM})^2}{B+S_\text{SM+BSM} + (\zeta_s S_\text{SM+BSM})^2 + (\zeta_b B)^2}.
\end{align}
Here, $B$ is the background rate, and $S_\text{SM}$ ($S_\text{SM+BSM}$) is the signal yield assuming SM couplings (nonzero $\delta a_\tau, \delta d_\tau$). At $\mathcal{L}=2~$nb$^{-1}$, we find $S_\text{SM} = 1280, B = 7.6$ for SR$1\ell$1T before binning in $p_\text{T}^\ell$; $S_\text{SM} = 520, B = 15$ for SR$1\ell$2T; $S_\text{SM} = 370, B = 4$ for SR$1\ell$3T. We denote the relative signal (background) systematic uncertainties by $\zeta_s$ ($\zeta_b$) and study $\zeta_s = \zeta_b \in [5\%, 10\%]$ as benchmarks. For simplicity, we assume identical $\zeta_s$ for all couplings, and combine the four SRs (SR$1\ell$1T has two $p_\text{T}^\ell$ bins) using $\chi^2 = \sum\chi^2_\text{SR}$ assuming uncorrelated systematics. We define the 68\% CL (95\% CL) regions as couplings satisfying $\chi^2 < 1$ ($\chi^2 < 3.84$). Appendix~\ref{sec:baseline} details cutflows for signals and backgrounds, and $\chi^2$ distributions.

\begin{figure}
    \centering
    \includegraphics[width=0.49\textwidth]{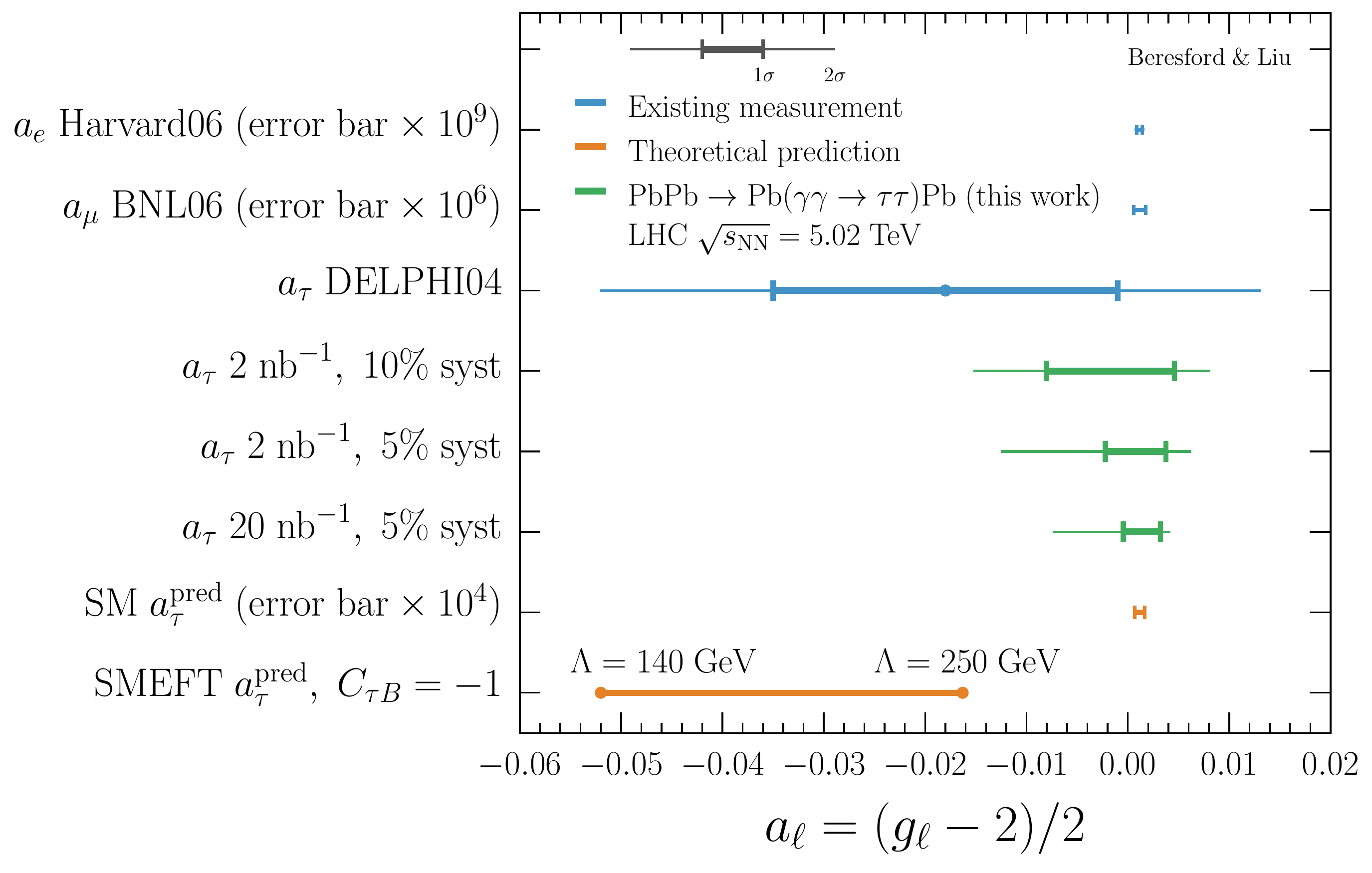}
    \caption{Summary of lepton anomalous magnetic moments $a_\ell = (g_\ell -2)/2$. Existing single-experiment measurements of $a_e$~\cite{PhysRevLett.97.030801}, $a_\mu$~\cite{Bennett:2006fi}, and $a_\tau$~\cite{Abdallah:2003xd} are in blue. Our benchmark projections (green) assume 2~nb$^{-1}$ and 20~nb$^{-1}$ for 5\% and 10\% systematic uncertainties. For visual clarity, we inflate $1\sigma$ error bars on $a_e$ ($a_\mu$) measurements by $10^9$ ($10^6$), and $10^4$ for the SM prediction $a_\tau^\text{pred}$ (orange)~\cite{Eidelman:2007sb}. Collider constraints have thick (thin) lines denoting 68\% CL, $1\sigma$ (95\% CL, $\sim2\sigma$). The SMEFT predictions~\cite{Escribano:1993pq,Grzadkowski:2010es} from Eq.~\eqref{eq:delta_a_d_tau_defn} with $C_{\tau B} = -1$ displays BSM scales $140 < \Lambda < 250$~GeV (thick orange).  }
    \label{fig:chi_sq}
\end{figure}

Figure~\ref{fig:chi_sq} summarizes our projected $a_\tau = a_{\tau, \,\mathrm{SM}}^\text{pred} + \delta a_\tau$ constraints (green) compared with existing measurements and predictions. Assuming the current dataset $\mathcal{L} = 2$~nb$^{-1}$ with 10\% systematics, we find $-0.0080 < a_\tau < 0.0046$ at 68\% CL, surpassing DELPHI precision~\cite{Abdallah:2003xd} (blue) by a factor of three. Negative values of $\delta a_\tau$ are more difficult to constrain given destructive interference. We estimate prospects assuming halved systematics giving $-0.0022 < a_\tau < 0.0037$ (68\% CL). A tenfold dataset increase for the High Luminosity LHC (HL-LHC) reduces this to $-0.00044 < a_\tau < 0.0032$ (68\% CL), an order of magnitude improvement beyond DELPHI. Importantly, these advances start constraining the sign of $a_\tau$ and becomes comparable to the predicted SM central value for the first time.

Such precision indirectly probes BSM physics. In nature, compositeness can induce large and negative magnetic moments e.g.\ the neutron~\cite{Tanabashi:2018oca}. As a benchmark, we fix $C_{\tau B} = -1, C_{\tau W} = 0, \delta d_\tau = 0$ in Eq.~\ref{eq:BSMLagrangian} to recast the DELPHI limit into a 95\%~CL exclusion of $\Lambda < 140$~GeV. The orange line in Fig.~\ref{fig:chi_sq} shows $140<\Lambda < 250$~GeV, where our 2~nb$^{-1}$, 10\% systematics proposal has $95\%$~CL sensitivity, surpassing DELPHI by 110~GeV. In suitable ultraviolet completions of SMEFT with composite leptons, one can interpret $\Lambda$ as the confinement scale of tau substructure~\cite{Silverman:1982ft}. Nonetheless, our analyses are highly model-independent and we defer sensitivity to other BSM scenarios for future work. It would be interesting to correlate $a_\tau$ with models that simultaneously explain tensions in $a_e$ and $a_\mu$~\cite{Dutta:2018fge,Davoudiasl:2018fbb,Bauer:2019gfk} or $B$-physics lepton universality tests~\cite{Aaij:2015yra,Abdesselam:2019dgh,Allanach:2015gkd,DiChiara:2017cjq,Biswas:2019twf}. 

Lepton electric dipole moments are highly suppressed in the SM, arising only at four-loop $|d_\tau^\text{pred}| \sim (m_\tau / m_e)|d_e^\text{pred}| \sim 10^{-33}~e$~cm~\cite{Pospelov:1991zt}. Additional CP violation in the lepton sector can enhance this, such as neutrino mixing~\cite{Ng:1995cs}, or other BSM physics parameterized by $\varphi$ in Eq.~\ref{eq:delta_a_d_tau_defn}. Our projected 95\% CL sensitivity on $d_\tau =  (e/m_\tau) \delta d_\tau $   is $|d_\tau| < 3.4 \times 10^{-17}~e$~cm, assuming $\delta a_\tau = 0$ with 2~nb$^{-1}$, 10\% systematics. This is an order of magnitude better than DELPHI $|d_\tau| < 3.7 \times 10^{-16}~e$~cm~\cite{Abdallah:2003xd} and competitive with Belle~\cite{Inami:2002ah}.

Our proposal opens numerous avenues for extension. Lowering lepton/track thresholds to increase statistics would enable more optimized differential or multivariate analyses. Recently, ATLAS considered tracks matched to lepton candidates failing quality requirements, allowing $p_\text{T}^\text{track}(e/\mu) > 1/2$~GeV~\cite{ATLAS-CONF-2019-014}.
Moreover the 500~MeV track threshold is conservative given $p_\text{T}^\text{track}>100$~MeV is successfully used in ATLAS~\cite{Aad:2019ock}. Reconstructing soft calorimeter clusters could enable hadron/electron identification, or using neutral pions to improve tau momentum resolution. Proposed timing detectors may offer more robust particle identification in ATLAS/CMS~\cite{CMSMIP:2296612} while ALICE already has such capabilities~\cite{Yu:2013dca}. Ultimate $a_\tau$ precision requires a coordinated worldwide program led by LHC efforts combined with proton--lead collisions at $\sqrt{s_\text{NN}} = 8.76$~TeV, Relativistic Heavy Ion Collider (RHIC), and lepton colliders. 

%\section{\label{sec:conclusion} Conclusion}

To summarize, we proposed a strategy of lepton plus track(s) analyses to surpass LEP constraints on tau electromagnetic moments using heavy ion data already recorded by the LHC. The clean photon collision events provide excellent opportunities to optimize low momentum reconstruction and control systematics further. We encourage LHC collaborations to open these cornerstone measurements and precision pathways to new physics.

%\begin{acknowledgements}
%\vspace{-2ex}

\emph{\textbf{Acknowledgements}}---We thank the hospitality of the LHC Forward and Diffractive Physics Workshop at CERN, where part of this work began. We are grateful to Luca Ambroz, Bill Balunas, Alan Barr, Mikkel Bj\o rn, Barak Gruberg, Lucian Harland-Lang, Simon Knapen, Santiago Paredes, Hannah Pullen, Hayden Smith, Beojan Stanislaus, Gabija \v{Z}emaityt\.{e} and Miha Zgubi\v{c} for helpful discussions. LB is supported by a Junior Research Fellowship at St John's College, Oxford. JL is supported by an STFC Postgraduate Studentship at Oxford, where this work started, and the Grainger Fellowship. 
%\end{acknowledgements}

\bibliography{bibs/intro.bib,bibs/pheno.bib,bibs/software.bib,./bibs/exp.bib,./bibs/theory.bib,./bibs/upc.bib}

\onecolumngrid
\appendix

\section{\label{sec:apndx_simulation}Simulation validation}

We present additional material to validate the technical implementation of our simulation setup models the intended physics effects within the scope of our work. This includes the photon flux we implemented in \textsc{MadGraph}~2.6.5~\cite{Alwall:2011uj,Alwall:2014hca}, and the interface with \textsc{SMEFTsim}~\cite{Brivio:2017btx} for BSM modifications and interference with the SM.

\begin{figure*}[h!]
    \centering
    \includegraphics[width=0.49\textwidth]{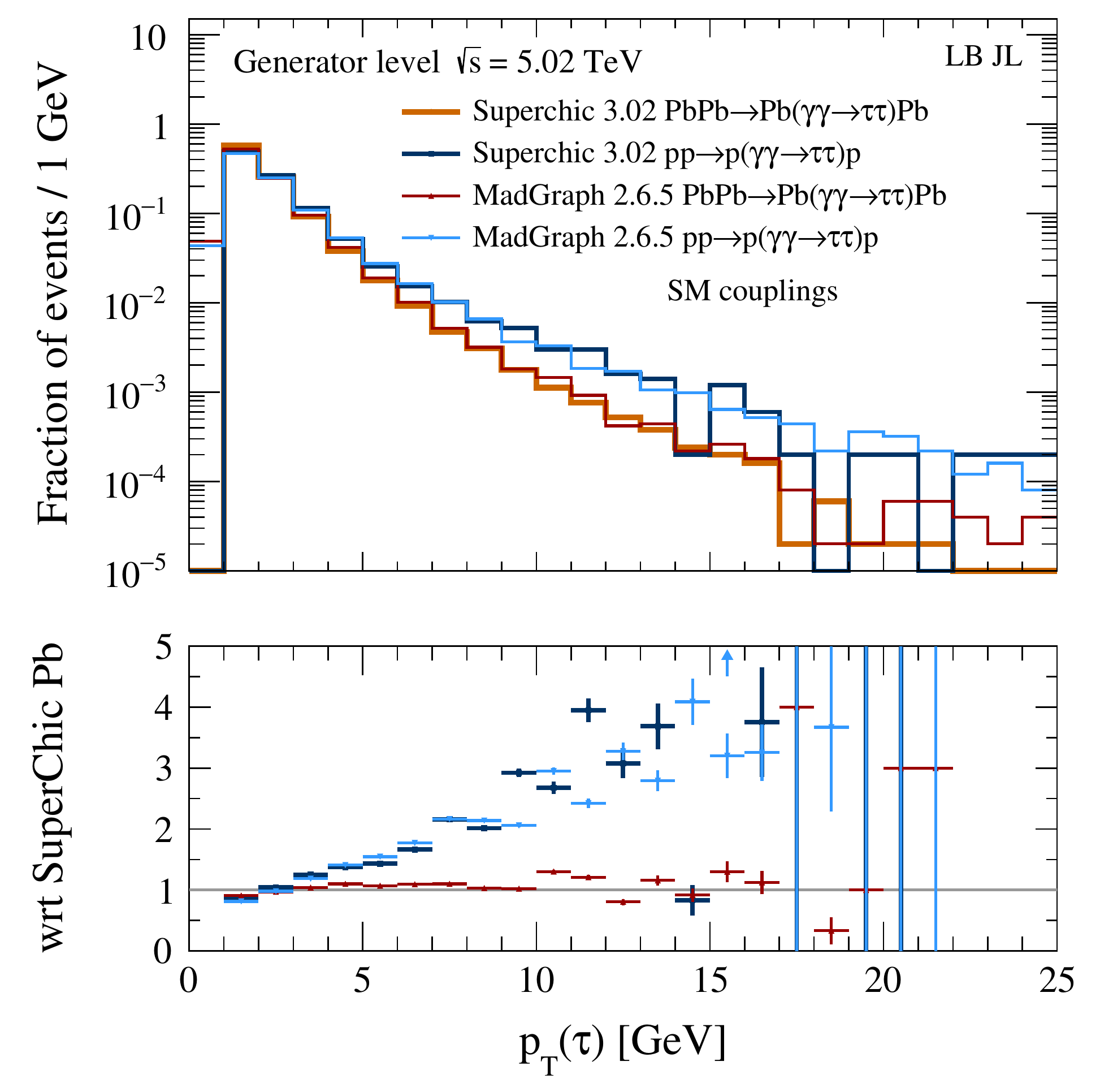}
    \caption{ Unit normalized generator level tau $p_\text{T}$ distributions for $\gamma\gamma\to \tau\tau$ using SM couplings. These are generated in \textsc{Superchic}~3.02, which includes a full treatment of nuclear effects for lead (Pb) ions (orange). Also shown is the corresponding sample with protons (dark blue). The \textsc{MadGraph}~2.6.5 samples uses a factorized photon flux prescription for protons (light blue) and our implementation of Pb ion flux (red). The ratio panel is with respect to the \textsc{Superchic} Pb ions sample.}
    \label{fig:MG_SC_simulation}
\end{figure*}

Figure~\ref{fig:MG_SC_simulation} displays generator level differential distributions of $p_\text{T}(\tau)$ for $\gamma\gamma\to\tau\tau$ considering various photon fluxes from protons and lead (Pb) beams. The distribution generated in \textsc{MadGraph} with Pb uses our custom implementation of Pb ion photon flux. We validate this with the corresponding distribution generated in \textsc{Superchic}~3.02~\cite{Harland-Lang:2018iur}. The latter includes a full treatment of nuclear effects that are neglected by the factorized prescription in \textsc{MadGraph}. These two distributions are in reasonable agreement for the scope of our work. Also shown are the corresponding distributions for proton beams. This illustrates that the impact of a nucleus with comparatively finite size is to soften the $p_\text{T}(\tau)$ spectrum compared to using proton beams.

\begin{figure*}[h!]
    \centering
    \includegraphics[width=0.49\textwidth]{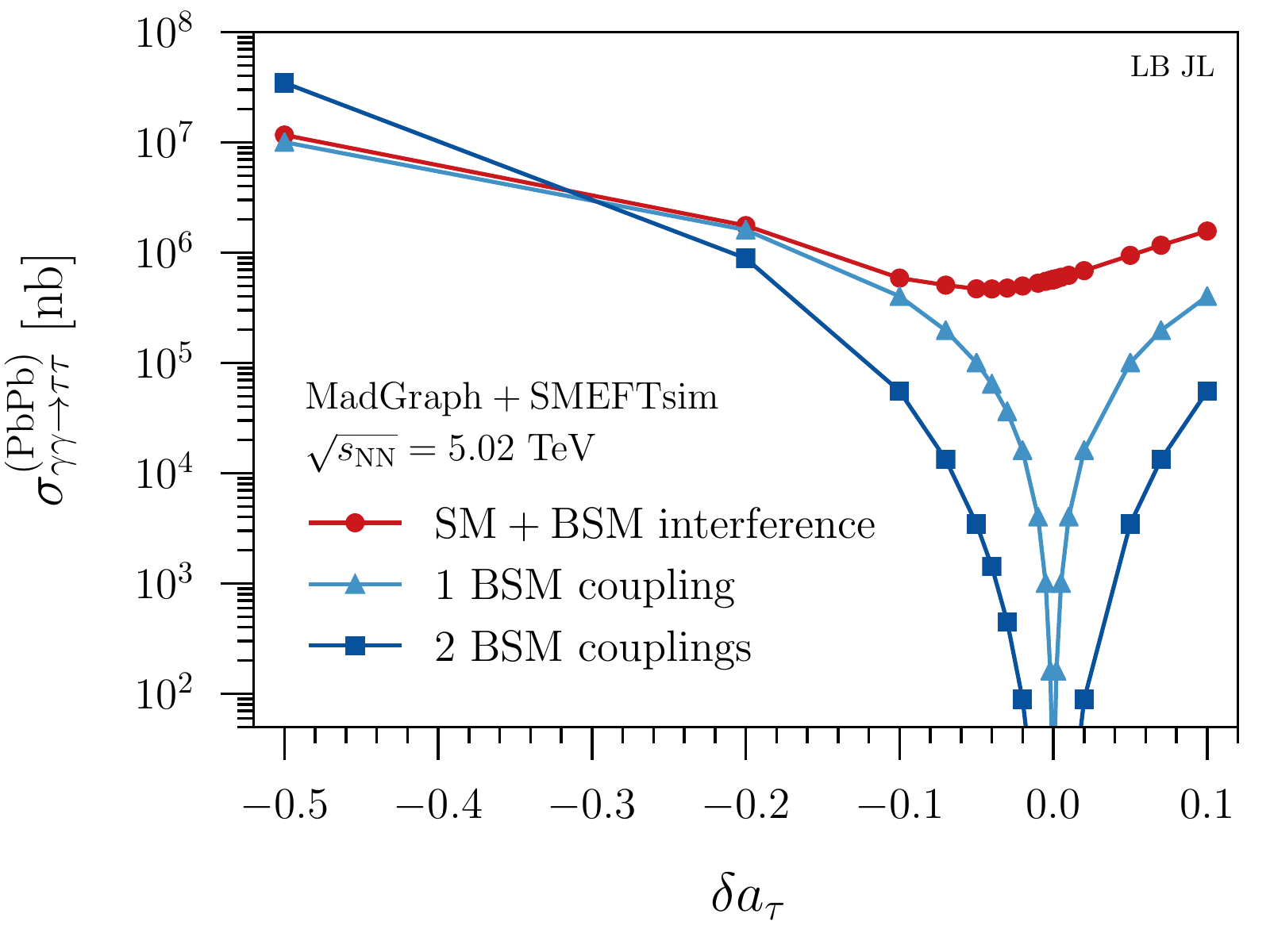}
    \includegraphics[width=0.49\textwidth]{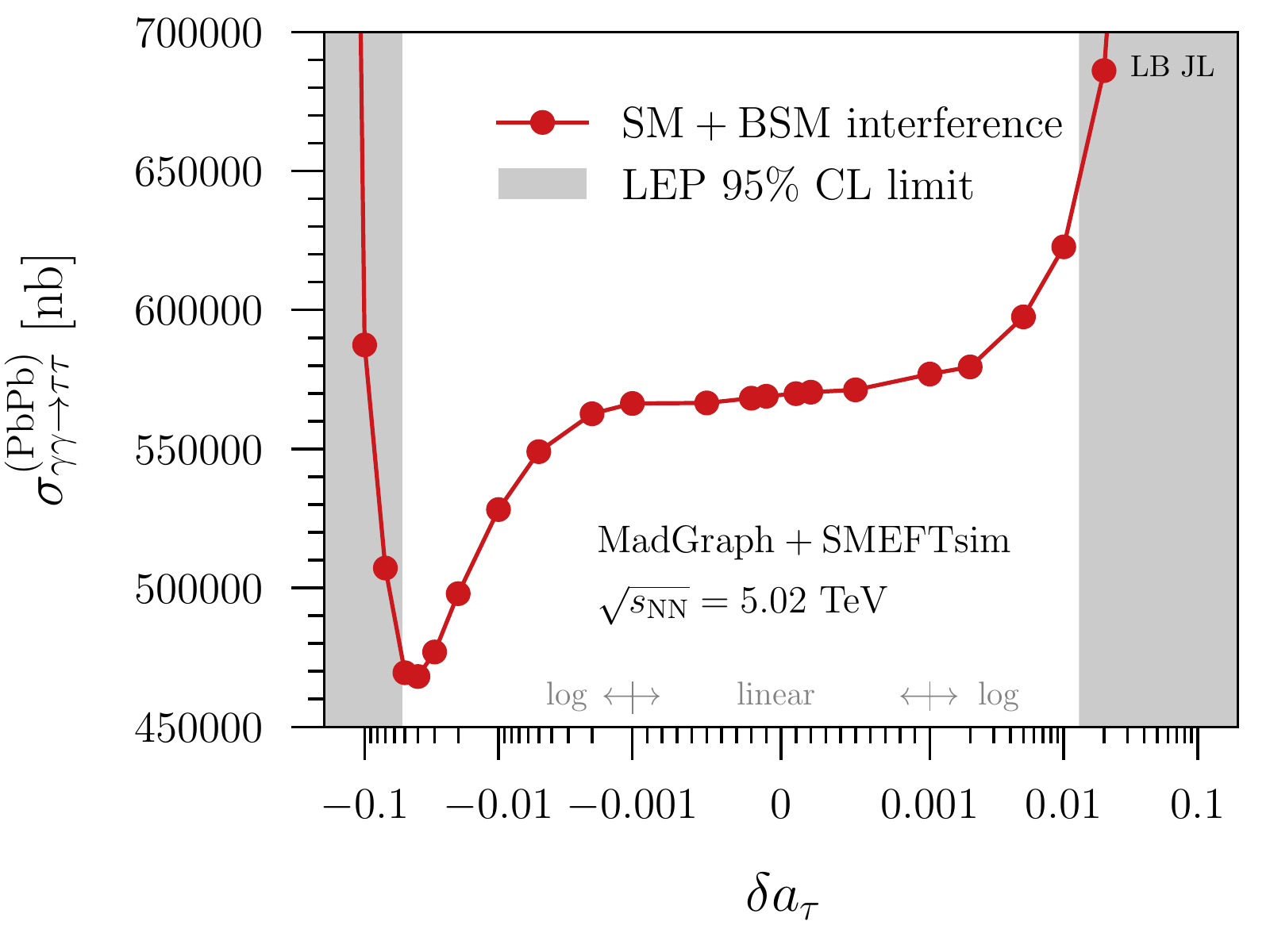}
    \caption{Generator level cross-sections for $\gamma\gamma \to \tau\tau$ sourced by our implementation of the Pb photon flux in \textsc{MadGraph}. This is interfaced with \textsc{SMEFTsim} for BSM coupling variations in $\delta a_\tau$ defined in Eq.~\ref{eq:delta_a_d_tau_defn} of the main text, fixing $\delta d_\tau = 0$ at $\sqrt{s_\text{NN}} = 5.02$~TeV. Left shows the contribution from only 1 BSM coupling (light blue triangles), 2 BSM couplings (dark blue squares), and their combined interference with the SM (red circles). The markers indicate the sampled points from $\delta a_\tau$. 
    Right zooms in to the $\delta a_\tau$ values near zero with gray regions denoting the 95\% CL exclusion by DELPHI, where the horizontal axis is linear scale for $\delta a_\tau \in [-0.001, 0.001]$ and logarithmic elsewhere. }
    \label{fig:aa2tautau_xsec}
\end{figure*}

Figure~\ref{fig:aa2tautau_xsec} shows the impact of the interference behavior on the inclusive cross-sections of $\sigma_{\gamma\gamma \to \tau\tau}^{(PbPb)}$ for coupling variations $\delta a_\tau$ using \textsc{SMEFTsim}. We account for the interference between SM and BSM $\gamma\gamma \to \tau\tau$ diagrams in the matrix element $\mathcal{M}$ squared
\begin{align}
|\mathcal{M}|^2 &= \left|\mathcal{M}_\text{SM} + \mathcal{M}_{\text{BSM}}^{(1)} + \mathcal{M}_{\text{BSM}}^{(2)}\right|^2\\
&=  \begin{gathered}
\includegraphics{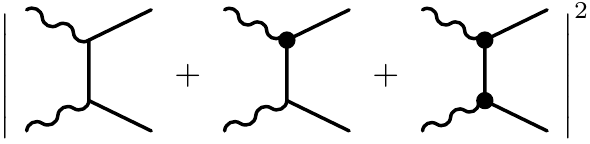}
\end{gathered}.
\end{align}
A BSM coupling is represented by $\bullet$ in the matrix element diagrams. Cross-sections featuring just the diagrams with only 1 BSM coupling (blue triangle) and only 2 BSM couplings (blue square) are shown in Fig.~\ref{fig:aa2tautau_xsec}, which correspond to the amplitudes $\mathcal{M}_\text{BSM}^{(1)}$ and $\mathcal{M}_\text{BSM}^{(2)}$ respectively. As $\delta a_\tau$ deviates from zero in the negative direction, $\sigma_{\gamma\gamma \to \tau\tau}^\text{(PbPb)}$ falls to a minimum at $\delta a_\tau \simeq -0.04$ due to destructive interference from $\mathcal{M}_\text{BSM}^{(1)}$. Then, the constructively interfering $\mathcal{M}_\text{BSM}^{(2)}$ term begins to dominate for more negative $\delta a_\tau$ values and $\sigma_{\gamma\gamma \to \tau\tau}^\text{(PbPb)}$ rises again.

\pagebreak
\section{\label{sec:baseline}Cutflows and $\chi^2$ distributions}

We provide technical material supporting the results presented in the main text. These include signal and background counts after sequentially applying kinematic requirements (cutflow), and $\chi^2$ distributions as functions of $\delta a_\tau$ and $\delta d_\tau$ used to derive the final constraints. 

\begin{table}[!ht]
\centering
\renewcommand{\arraystretch}{1.06}
\resizebox{\linewidth}{!}{
  \begin{tabular*}{\textwidth}{@{\extracolsep{\fill}}lrrrrrrrrrr}
  \toprule
  Requirement  
 & $\tau\tau$ $(0,0)$
 & $\tau\tau$ $(0.005,0)$ 
 & $\tau\tau$ $(-0.01,0)$
 & $\mu\mu$
 & $ee$
 & $bb$
 & $cc$
 & $ss$
 & $uu$
 & $dd$\\
  \midrule
  \multicolumn{2}{l}{1 lepton + 1 track analysis (SR$1\ell$1T)}\\
  \midrule
  $\sigma \times \mathcal{L}$ & $1139800$  & $1195060$  & $1056400$  & $844080$  & $844080$  & $2999$  & $604080$  & $37754$  & $604080$  & $37754$ \\
  $\sigma  \times \mathcal{L} \times \epsilon_\text{filter}$ & $241140$  & $253920$  & $226300$  & $844080$  & $844080$  & $2999$  & $604080$  & $37754$  & $604080$  & $37754$ \\
    $1\ell$ plus 1 track & $20492.2$  & $21619.3$  & $19348.4$  & $263443$  & $3299.3$  & $5.4$  & $2905.0$  & $0.3$  & $5.4$  & $0.2$ \\
    $p_\text{T}^{e/\mu} > 4.5/3~\text{GeV}$, $|\eta^{e/\mu}| < 2.5/2.4$ & $3659.9$  & $3882.7$  & $3582.8$  & $79043$  & $3118.9$  & $1.1$  & $4.8$  & $0.0$  & $0.0$  & $0.0$ \\
    2 tracks, $p_\text{T}^{\text{trk}} > 0.5~\text{GeV}$, $|\eta^{\text{trk}}| <$ 2.5 & $3324.5$  & $3535.9$  & $3256.9$  & $78973$  & $3117.8$  & $1.0$  & $3.0$  & $0.0$  & $0.0$  & $0.0$ \\
    $|\Delta\phi(\ell,\text{trk})| < 3$ & $1519.7$  & $1605.7$  & $1468.3$  & $0.9$  & $5.3$  & $0.7$  & $1.8$  & $0.0$  & $0.0$  & $0.0$ \\
    $m_{\ell,\text{trk}} \not\in \{[3, 3.2], [9, 11]\}$~GeV  & $1275.1$  & $1353.6$  & $1242.3$  & $0.9$  & $5.3$  & $0.2$  & $1.2$  & $0.0$  & $0.0$  & $0.0$ \\
    \midrule
    $p_\text{T}^{\ell} \leq 6.0~\rm{GeV}$ & $1197.7$  & $1262.3$  & $1154.7$  & $0.9$  & $0.0$  & $0.2$  & $1.2$  & $0.0$  & $0.0$  & $0.0$ \\
    \midrule
     $p_\text{T}^{\ell} > 6.0~\rm{GeV}$ & $77.3$  & $91.3$  & $87.6$  & $0.0$  & $5.3$  & $0.0$  & $0.0$  & $0.0$  & $0.0$  & $0.0$ \\
    \midrule\midrule
    \multicolumn{2}{l}{1 lepton + multitrack analysis (SR$1\ell$2/3T)}\\
    \midrule
    $\sigma \times \mathcal{L}$ & $1139800$  & $1195060$  & $1056400$  & $844080$  & $844080$  & $2999$  & $604080$  & $37754$  & $604080$  & $37754$ \\
    $\sigma  \times \mathcal{L} \times \epsilon_\text{filter}$ & $241140$  & $253920$  & $226300$  & $844080$  & $844080$  & $2999$  & $604080$  & $37754$  & $604080$  & $37754$ \\
    $1\ell$ plus 2 or 3 tracks & $5945.1$  & $6260.1$  & $5572.2$  & $33.8$  & $23.2$  & $43.8$  & $8056.6$  & $5.4$  & $132.9$  & $6.8$ \\
    $p_\text{T}^{e/\mu} > 4.5/3~\text{GeV}$, $|\eta^{e/\mu}| < 2.5/2.4$  & $1010.0$  & $1073.3$  & $978.6$  & $12.2$  & $4.2$  & $1.8$  & $13.3$  & $0.0$  & $0.0$  & $0.0$ \\
    \midrule
    3 tracks, $p_\text{T}^{\text{trk}} > 0.5~\text{GeV}$, $|\eta|^{\text{trk}} <$ 2.5  & $519.9$  & $548.1$  & $485.8$  & $5.6$  & $4.2$  & $0.8$  & $4.8$  & $0.0$  & $0.0$  & $0.0$ \\
    \midrule
    4 tracks, $p_\text{T}^{\text{trk}} > 0.5~\text{GeV}$, $|\eta|^{\text{trk}} <$ 2.5 & $370.5$  & $398.3$  & $381.1$  & $0.0$  & $0.0$  & $0.4$  & $3.6$  & $0.0$  & $0.0$  & $0.0$ \\
  \bottomrule
  \end{tabular*}
  }
\caption{Cutflow of yields after each requirement applied sequentially, normalized to $\mathcal{L} = 2$~nb$^{-1}$ for the different analyses. For the $\gamma\gamma\to\tau\tau$ signal processes, we show these for benchmark points with parameter values labeled by $(\delta a_\tau, \delta d_\tau)$ displayed in the column header. Backgrounds are shown for various dilepton $\mu\mu, ee$ and diquark where the letters denote the flavor. The initial value in each cutflow is the cross-section $\sigma$ times luminosity $\mathcal{L}$, followed by the efficiency $\epsilon_\text{filter}$ of the filter applied at generator level to the $\gamma\gamma\to\tau\tau$ samples.
}
\label{tab:cutflow_combined}
\end{table}

Table~\ref{tab:cutflow_combined} presents the set of cutflows for the different analyses, sequentially displaying the yields normalized to 2~nb$^{-1}$ after each signal region requirement. Three benchmark signals are shown for the $\gamma\gamma\to\tau\tau$ samples at the SM values $(\delta a_\tau, \delta d_\tau) = (0, 0)$ and for values near the threshold of 68\% CL sensitivity $(\delta a_\tau, \delta d_\tau) \in \{(0.005, 0), (-0.01, 0)\}$.

\begin{figure*}[!ht]
    \centering
    \includegraphics[width=0.49\textwidth]{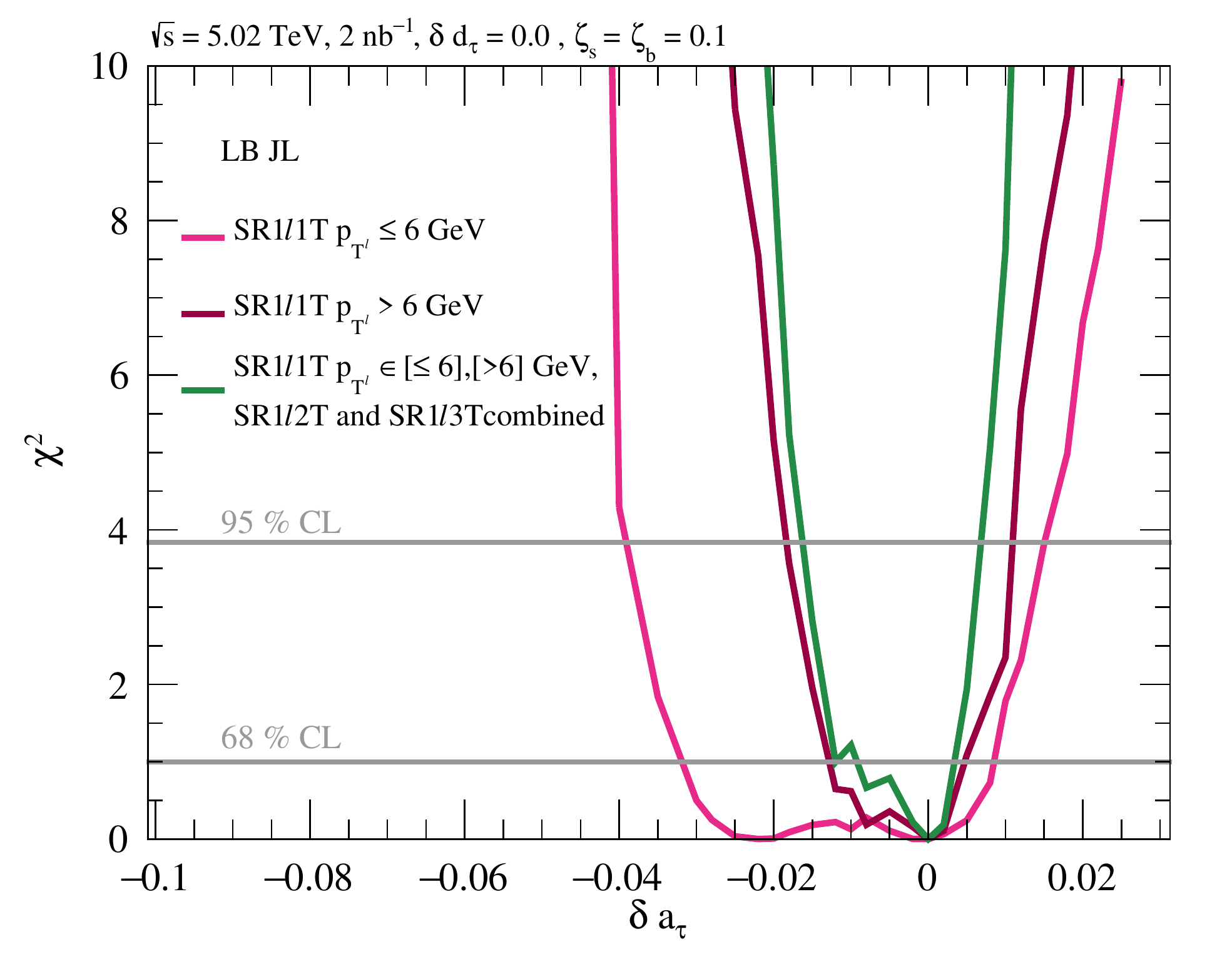}
    \includegraphics[width=0.49\textwidth]{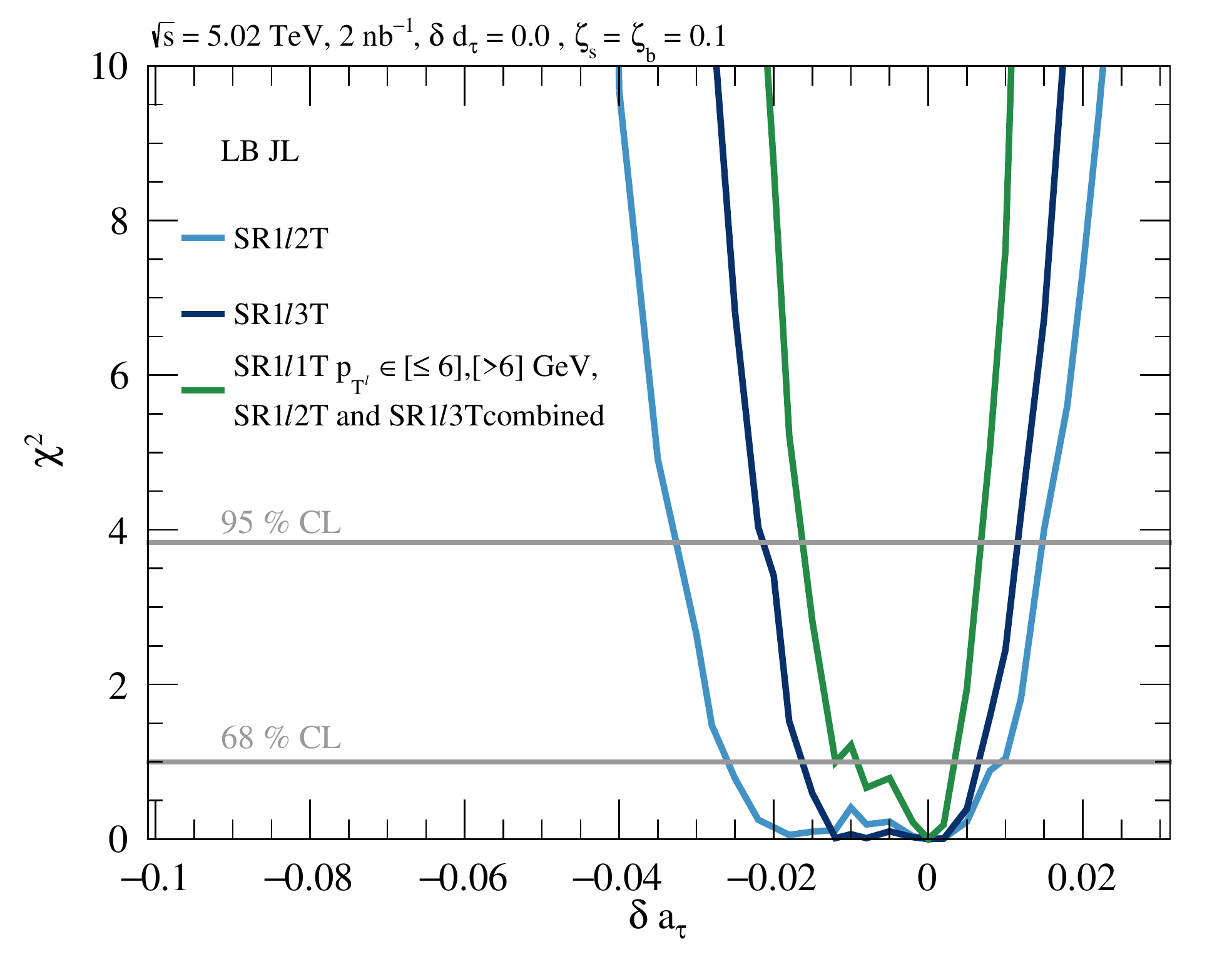}
    \includegraphics[width=0.49\textwidth]{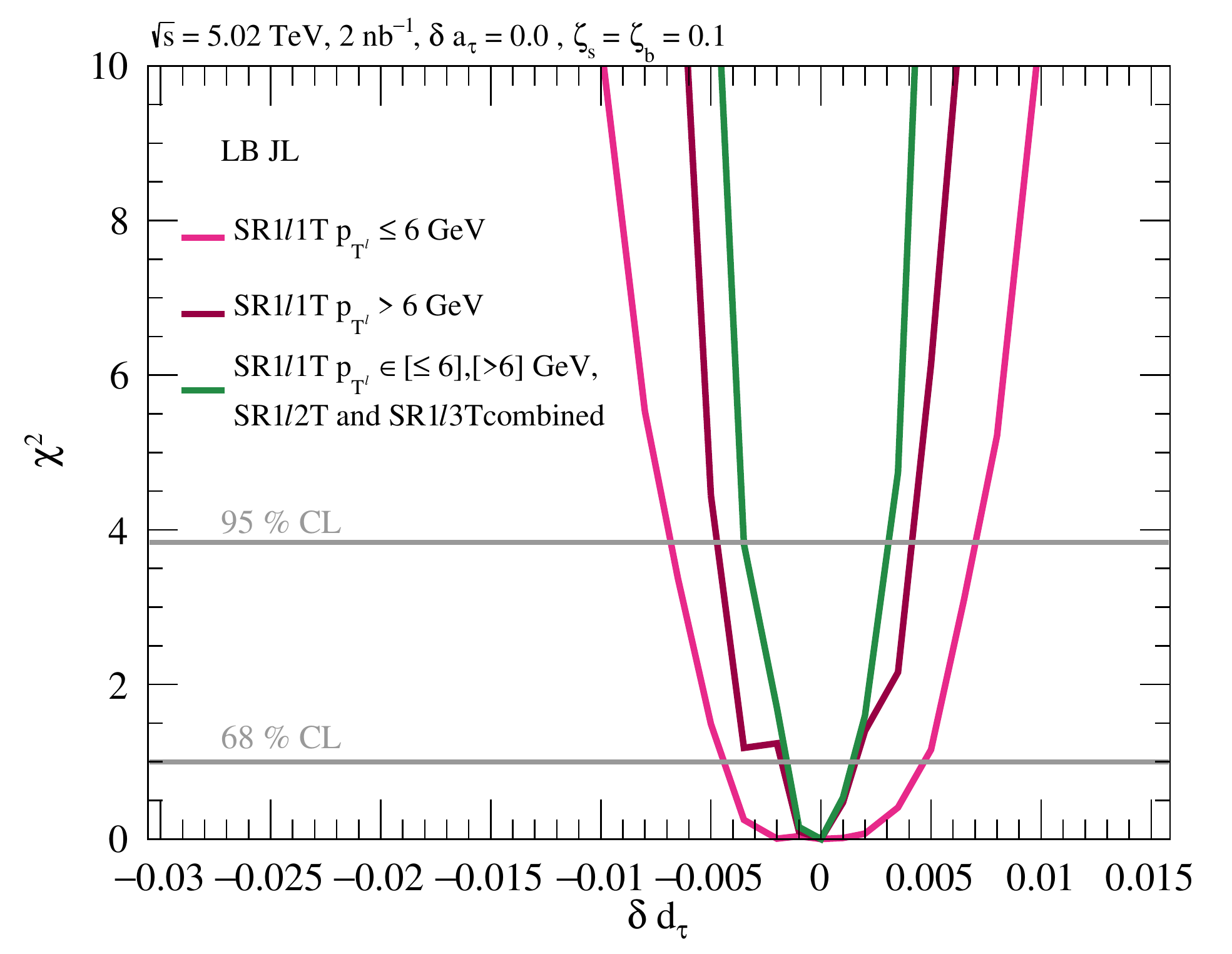}
    \includegraphics[width=0.49\textwidth]{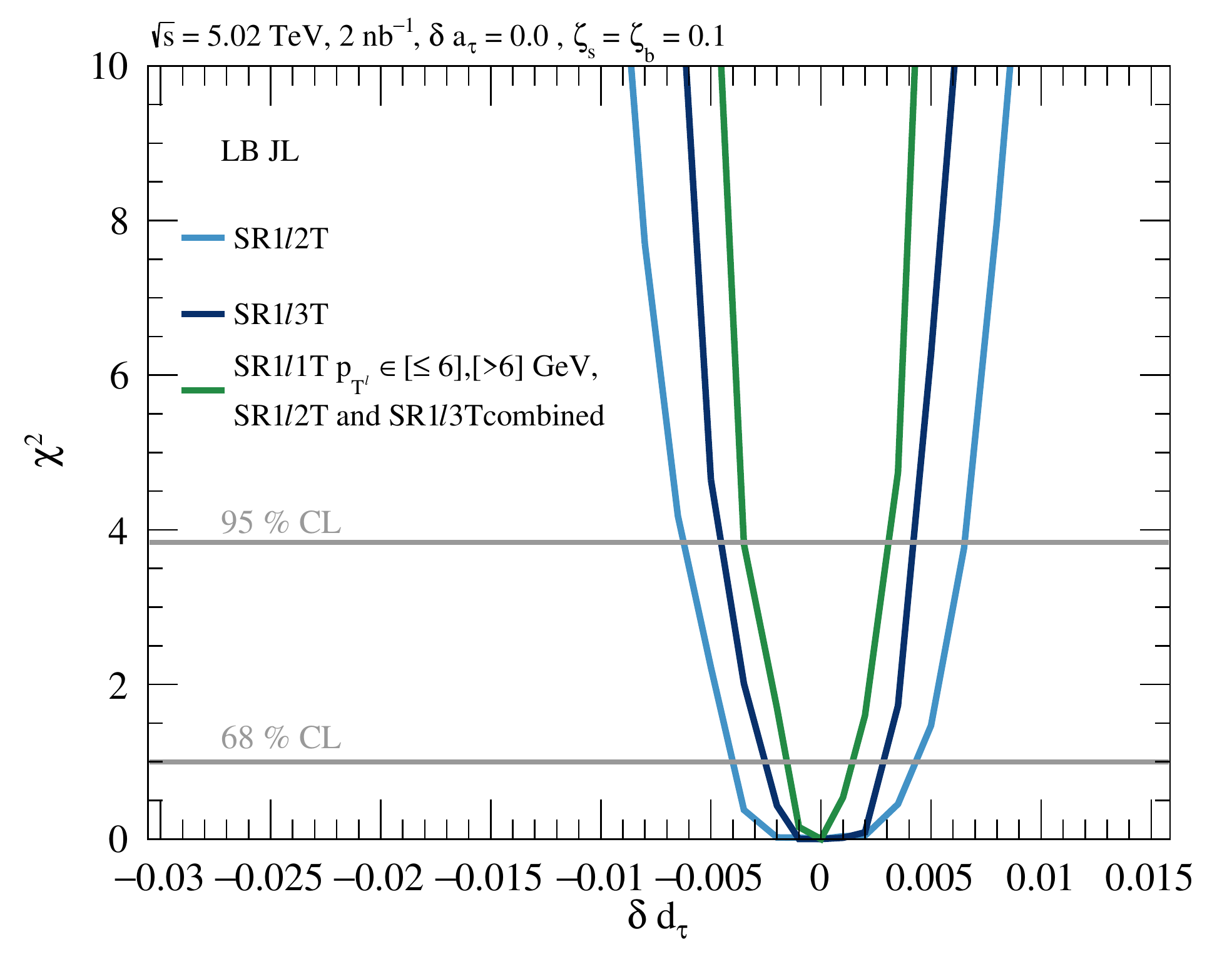}
    \caption{\label{fig:chiSq_distribution_separateSRs}
    The $\chi^2$ distributions as a function of $\delta a_{\tau}$ assuming $\delta d_{\tau}=0$ (upper), and $\delta a_{\tau}$ assuming $\delta d_{\tau}=0$ (lower) are displayed for 10\% systematics at $\mathcal{L} = 2$~nb$^{-1}$.   
    Left shows the results from the SR$1\ell$1T regions and the impact of binning in $p_\text{T}^\ell$.
    Right shows the results from the SR$1\ell$2/3T regions. The four signal region combined $\chi^2$ is shown by the green line for reference. The gray horizontal lines correspond to 68\% CL ($\chi^2$=1) and 95\% CL ($\chi^2$=3.84) 
    }
\end{figure*}

Figure~\ref{fig:chiSq_distribution_separateSRs} shows the $\chi^2$ distributions as a function of $\delta a_\tau$ and $\delta d_\tau$ assuming the other is zero for separate signal regions. These are shown assuming 10\% systematics, 2~nb$^{-1}$ to allow comparison of constraining power between the different analyses presented in the main text. 

\begin{figure*}[!ht]
    \centering
    \includegraphics[width=0.33\textwidth]{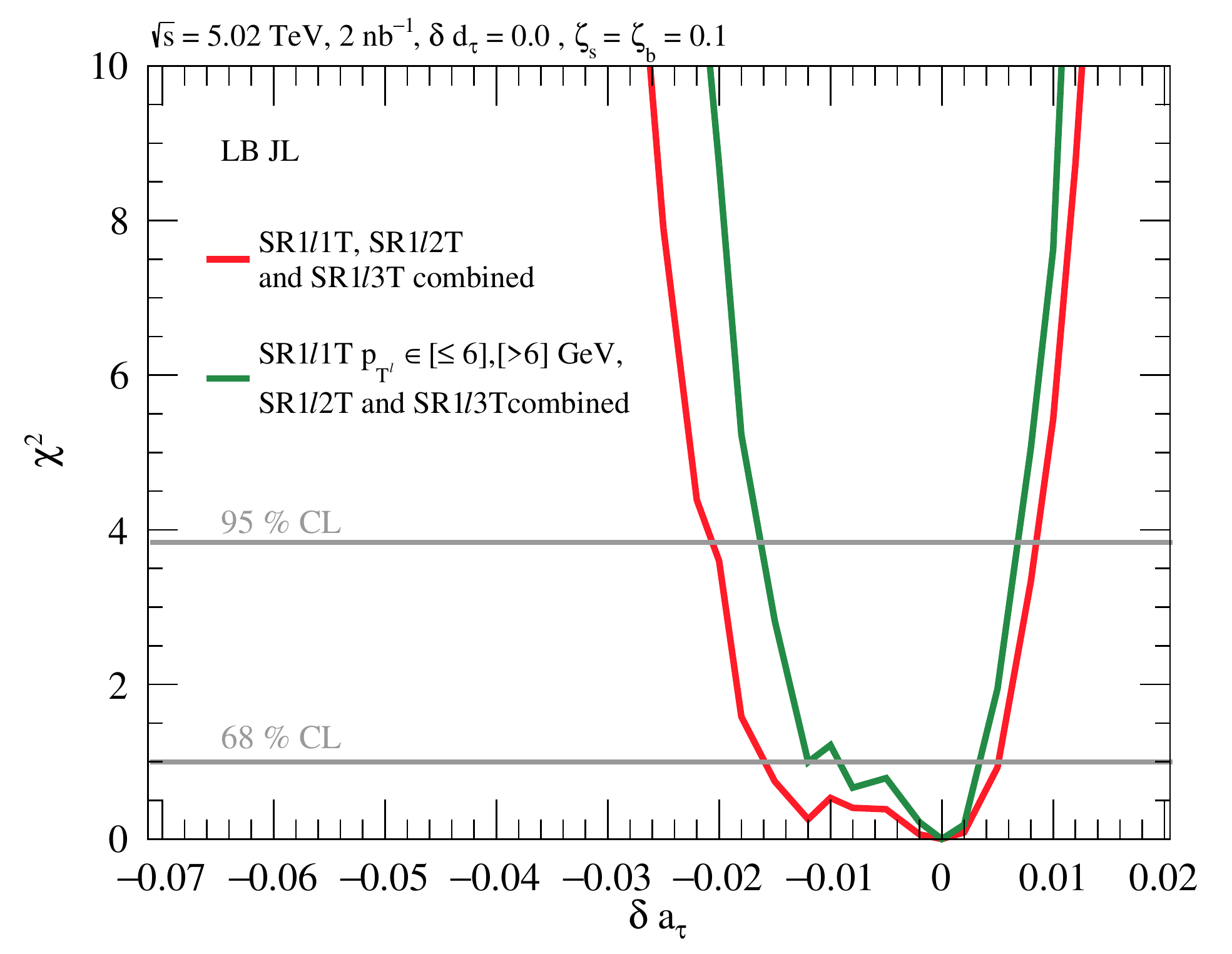}%
    \includegraphics[width=0.33\textwidth]{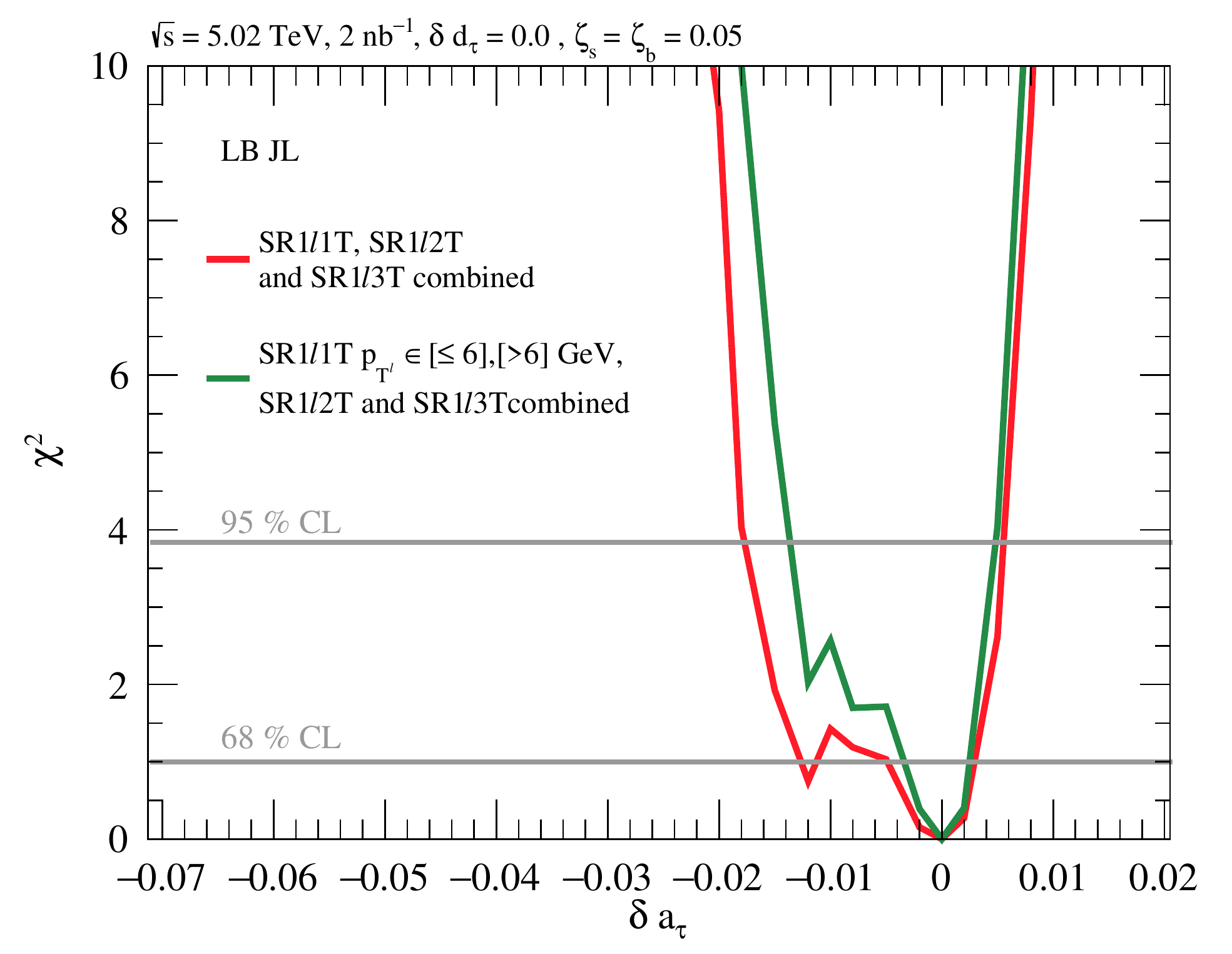}%
    \includegraphics[width=0.33\textwidth]{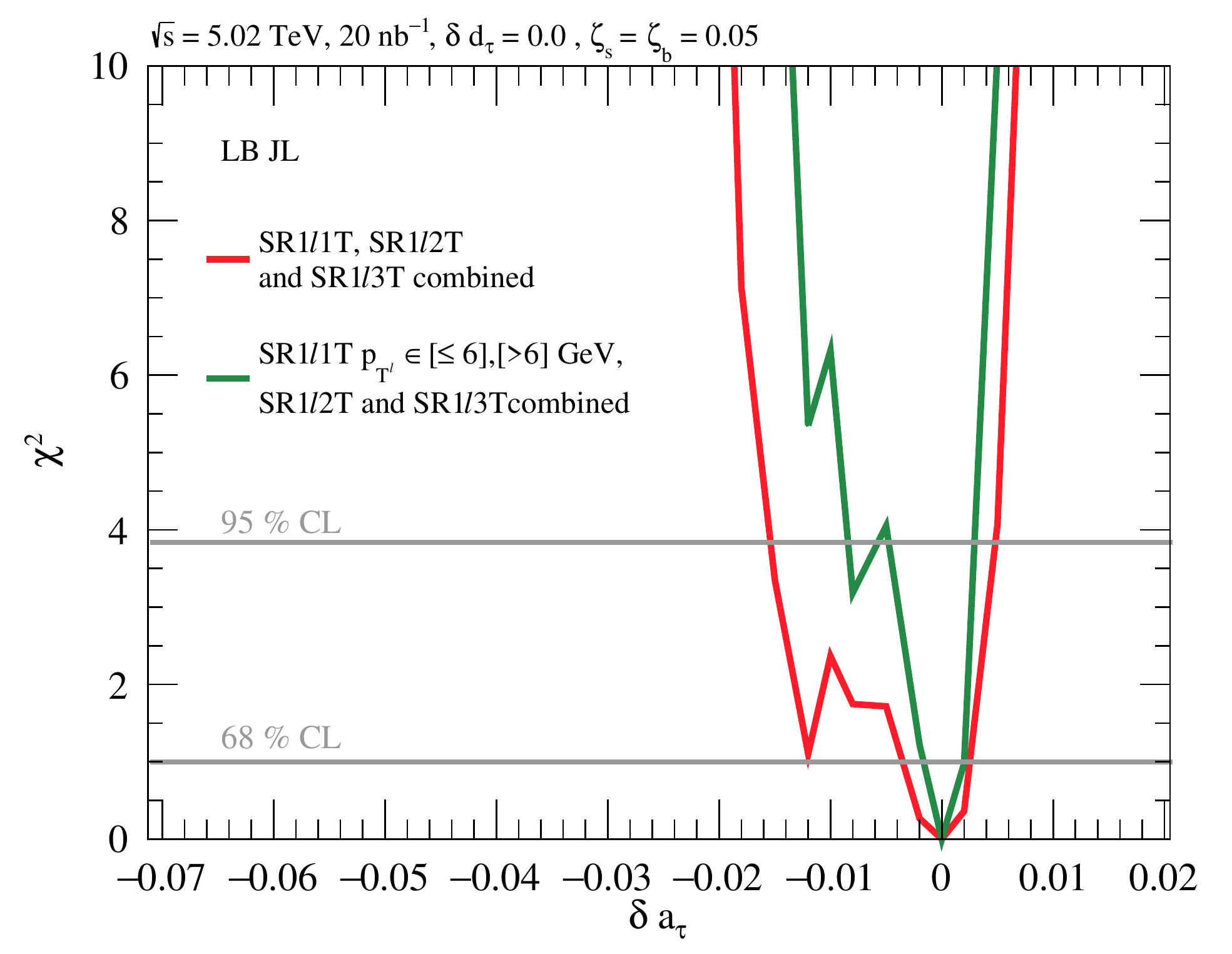}\\
    \includegraphics[width=0.33\textwidth]{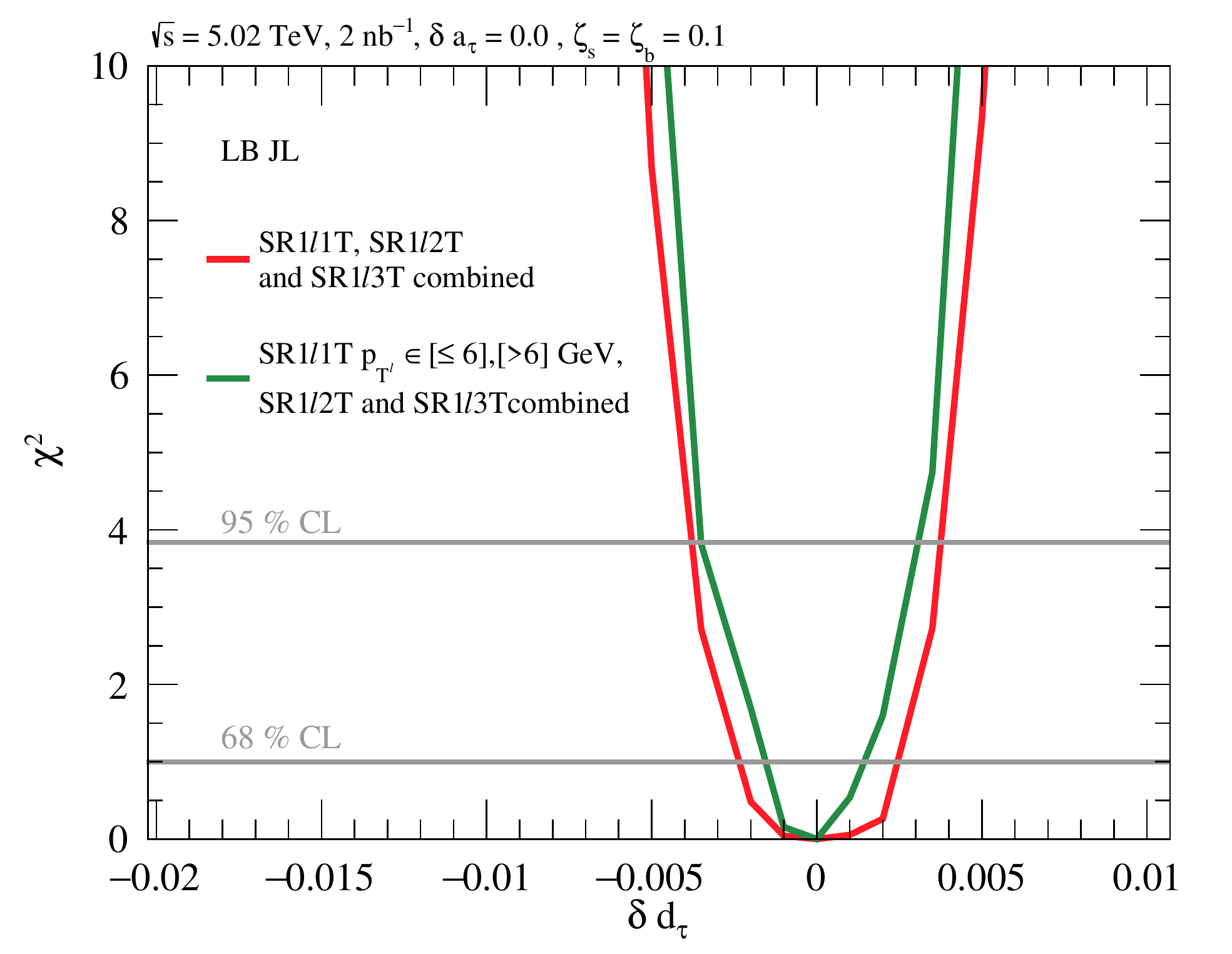}%
    \includegraphics[width=0.33\textwidth]{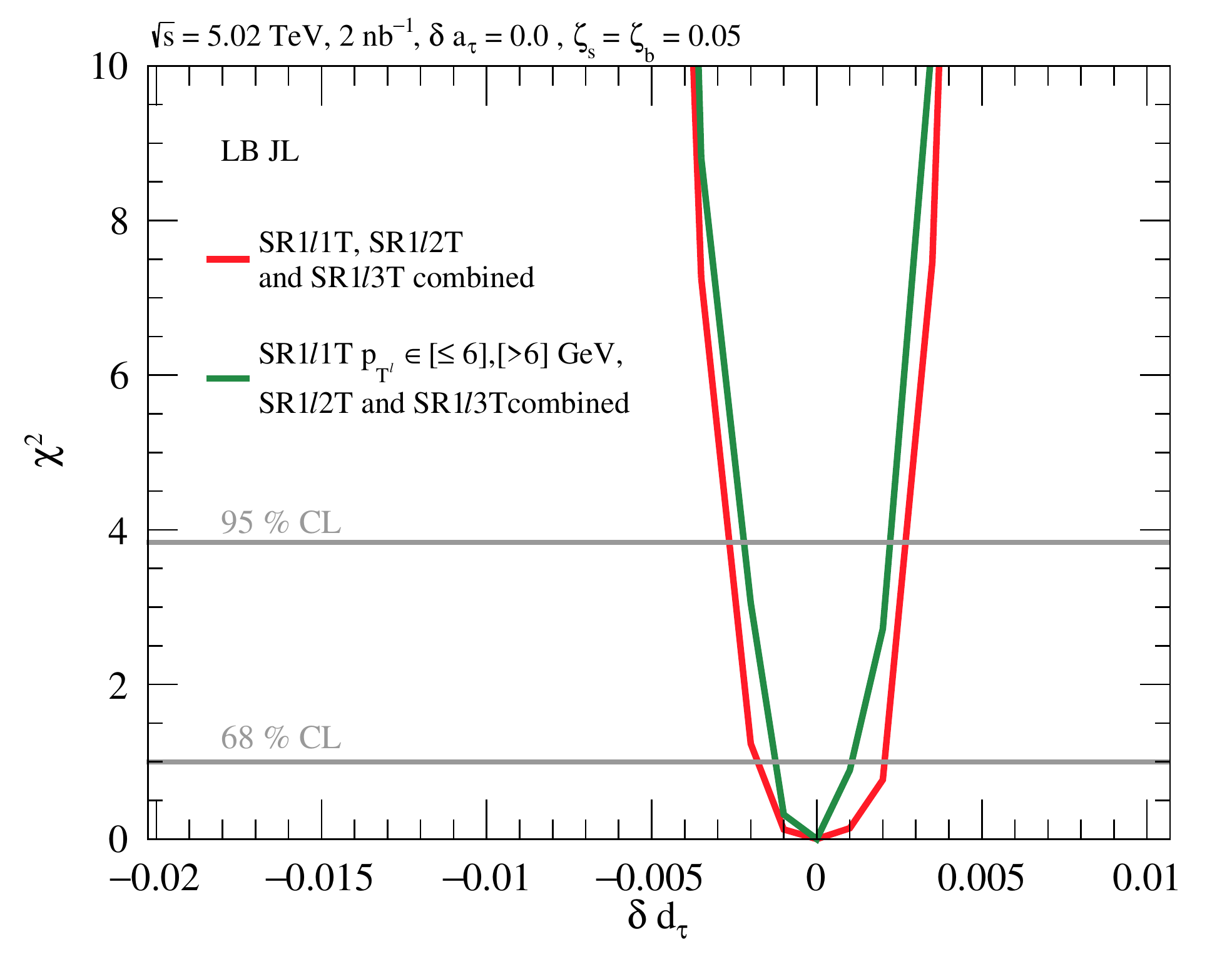}%
    \includegraphics[width=0.33\textwidth]{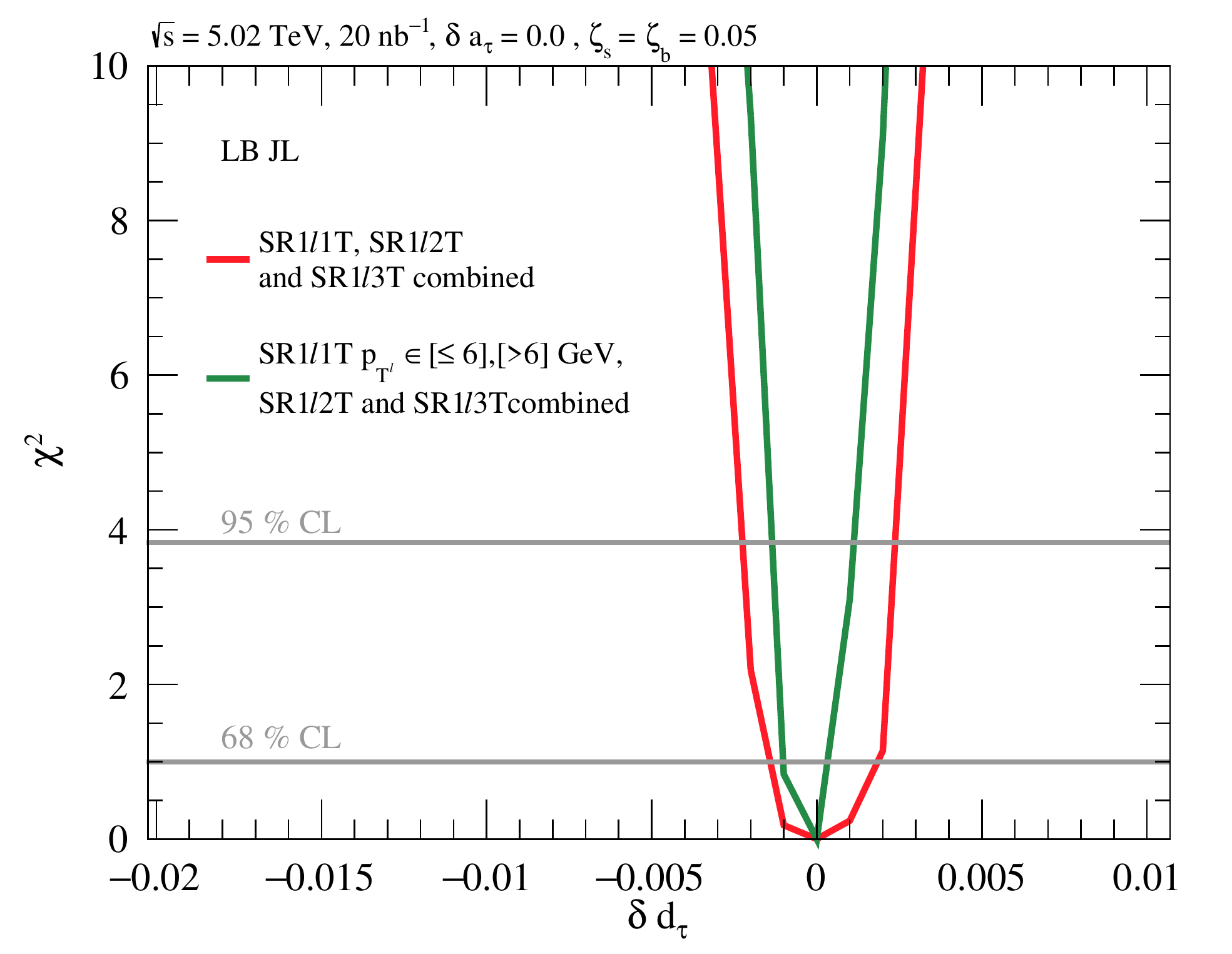}
    \caption{\label{fig:chiSq_distributions_combined} 
    The $\chi^2$ distributions as a function of $\delta a_{\tau}$ assuming $\delta d_{\tau}=0$ (upper), and $\delta a_{\tau}$ assuming $\delta d_{\tau}=0$ (lower). These are displayed for 10\% (left), 5\% (centre) systematics at $\mathcal{L} = 2$~nb$^{-1}$, and 5\% systematics result extrapolated to $\mathcal{L} = 20$~nb$^{-1}$ (right).
    The combined $\chi^2$ for all three track SRs is shown by the red line, while the impact of dividing SR$1\ell$1T into two orthogonal $p_\text{T}^\ell$ bins is shown by the green line. The gray horizontal lines correspond to 68\% CL ($\chi^2$=1) and 95\% CL ($\chi^2$=3.84) 
    }
\end{figure*}

Figure~\ref{fig:chiSq_distributions_combined} displays the combined $\chi^2 = \sum_i \chi_i^2$ distributions. The combined $\chi^2$ distributions are shown for 10\% systematics at 2~nb$^{-1}$ together with prospects using 5\% systematics and extrapolation to 20~nb$^{-1}$. The red lines show the results from combining the three track SRs. The final combined $\chi^2$ for the results in the main text take the green lines, which combine all four signal regions (SR$1\ell$1T is divided into two orthogonal $p_\text{T}^\ell$ bins). The final 68\% CL and 95\% CL intervals are defined by where the $\chi^2$ distributions intersect with $\chi^2 = 1$ and $\chi^2 = 3.84$ respectively. 

\end{document}